%% file: msEG.tex
\documentclass[12pt]{article}
\usepackage{etex}
\usepackage{amsfonts}
\usepackage{latexsym}
\usepackage{natbib}
\usepackage{pst-pdf, pst-all, amsthm, amssymb, latexsym}
\usepackage{amsmath}
\usepackage{comment}
\usepackage{listings}
\usepackage{pstricks,pstricks-add} 
\usepackage{booktabs}
\usepackage{rotating}
\usepackage{multirow, tabularx}
\usepackage{graphicx}
\usepackage{lscape}
\usepackage{abstract}
\usepackage{lipsum} 
\usepackage{ctable}
\def\sym#1{\ifmmode^{#1}\else\(^{#1}\)\fi}
\usepackage{caption}
\usepackage{subfloat}
\usepackage{titlesec}
\usepackage{blindtext}
\usepackage{rotating}

\usepackage{caption}
\usepackage{subcaption}
\usepackage{mathtools}

\usepackage{pgfsys}

\makeatletter
  \newcommand\footnotesizes{\@setfontsize\footnotesizes{9.6pt}{6}}
\makeatother
\usepackage{graphicx}

\usepackage{epigraph}
\AtBeginDocument{%
  \addtolength\abovedisplayskip{-0.15\baselineskip}%
  \addtolength\belowdisplayskip{-0.15\baselineskip}%
  }
  \usepackage{lmodern}
  \usepackage{subcaption}

\makeatother
\usepackage{indentfirst}

\usepackage{enumerate}

\usepackage{hyperref}
\hypersetup{backref,  
pdfpagemode=FullScreen,  
colorlinks=true, citecolor=blue!50!black}
\usepackage{graphicx}

\renewcommand{\thetable}{\Roman{table}}
\renewcommand{\thefigure}{\Roman{figure}}

\usepackage{titlesec}

\titleformat{\paragraph}
{\normalfont\normalsize\bfseries}{\theparagraph}{1em}{}
\titlespacing*{\paragraph}
{0pt}{3.25ex plus 1ex minus .2ex}{1.5ex plus .2ex}

\usepackage{fancyhdr}

\usepackage{scalerel}
\newcommand\vfrac[2]{\ThisStyle{%
  \setbox0=\hbox{$\SavedStyle#1#2$}%
  \setbox2=\hbox{$\SavedStyle X$}%
  \ifdim\ht0>\ht2\setlength{\ht0}{\ht2}\fi%
  #1\mathord{\stretchto{\raisebox{2.3\LMpt}{$\SavedStyle/$}}{\ht0}}#2}}

\titleformat{\subsubsection}
  {\normalfont\fontsize{12}{17}\selectfont}{\thesubsubsection}{1em}{}
 
\usepackage{enumitem}

\usepackage{cleveref}
\usepackage{subcaption}

\usepackage{float}

\begin{document}

\title{Ideological Ambiguity and Political Spectrum}
\author{
 Hector Galindo-Silva\\
  \texttt{galindoh@javeriana.edu.co}\thanks{Department of Economics, Pontificia Universidad Javeriana.  I thank participants at several seminars, as well as the  editor and two anonymous referees for very helpful remarks and suggestions. Any remaining errors are my own.}}
\maketitle

\begin{abstract}
This study examines the relationship between ambiguity and the ideological positioning of political parties across the political spectrum. We identify a strong non-monotonic (inverted U-shaped) relationship between party ideology and ambiguity within a sample of 202 European political parties. This pattern is observed across all ideological dimensions covered in the data.  To explain this pattern, we argue that centrist parties are perceived as less risky by voters compared to extremist parties, giving them an advantage in employing ambiguity to attract more voters at a lower cost. We support our explanation with additional evidence from electoral outcomes and economic indicators in the respective party countries. 

\end{abstract}

\newpage


\section{Introduction}

Political parties often adopt ambiguous and inconsistent positions, a phenomenon that spans various countries, party systems, and time periods. This practice can limit voters' knowledge of the policies potential leaders intend to implement if elected, thus posing a considerable challenge to democracy. 

The primary objective of this study is to delve into the factors driving such electoral behavior, with a specific focus on a relatively intuitive yet insufficiently explored aspect: the degree of extremism or centrism inherent in political party ideologies.  We specifically investigate the relationship between ambiguity and the ideological positioning of political parties across the political spectrum.

Our main finding is the identification of a non-monotonic (inverted U-shaped) relationship between party ideology and ideological blurriness (or ambiguity)\footnote{In this study, ambiguity is defined as a deliberate strategy employed by political parties to avoid taking a clear position on a particular issue, effectively making it synonymous with strategic ambiguity, position-blurring, issue clarity, or ideological clarity. While there are some relevant differences between these terms (for instance, see the crucial distinction between ambiguity and vagueness highlighted by \citealp{Praprotniketalt2023}, or the important variations among avoidance, ambiguity, and alternation as studied by \citealp{Koedam2021}), for the purposes of this study, we believe that the general definition provided earlier is sufficient, given the data used. In addition, note that we assume ambiguity is strategic, in line with existing literature  \cite[e.g.][]{SomerTopcu2015, BrauningerGiger2018}.} within a sample of 202 European political parties. Specifically, our results indicate that political parties with a centrist ideology tend to possess a more blurred ideology. This relationship holds true for each ideological dimension covered in the available data. Furthermore, the relationship remains robust across different econometric specifications, providing strong evidence of its existence not only within countries but also within parties. While this study is descriptive in nature and a perfect identification strategy is challenging given the research question's nature, the broad range of correlational results, along with some instrumental variable (IV) estimates, provides support for the interpretation that the ideological position influences the degree of blurriness.

We also offer an  explanation for this pattern and present additional empirical evidence to support it. Our explanation builds upon \cite{Glazer1990} but expands on it. It is based on the notion that centrist political parties are perceived as less risky by voters compared to extremist parties. One possible reason is that policies proposed by extremist parties are often untested and unconventional, whereas centrist parties often propose more established policies. If the median voter is risk-averse, this suggests that centrist parties enjoy an ex ante advantage over extremist parties. Consequently, if ambiguity facilitates parties in aligning with the median voter, centrist parties have stronger incentives than extremist parties to strategically utilize ambiguity for electoral success. 

To strengthen the plausibility of our explanation, we analyze additional data on electoral outcomes and economic indicators from the respective countries of the studied parties. Our analysis reveals  that in countries where extremist parties have recently held power and experienced significant economic fluctuations, centrist parties not in government tend to adopt more ambiguous positions. This finding aligns with our theoretical framework, which suggests that incumbent extremist parties in such contexts should be perceived as riskier, thereby reducing the potential benefits of ambiguity for them. 

Furthermore,  we explore the empirical plausibility of two alternative explanations for the observed non-monotonic inverted U-shaped relationship between ideology and political position. Firstly, we examine whether this relationship could be attributed to the presence of single-issue parties that adopt centrist and ambiguous policies to attract a broader voter base, given that those policies are not their primary focus (e.g., \citealp{Rovny2012, SomerTopcu2015}). Secondly, we investigate whether this relationship can be explained by the greater interest that centrist parties may have in having more flexibility while in office to choose policies that better deal with new information not available during elections (e.g., \citealp{AragonesNeeman2000, Kartiketalt2017}). Our empirical analyses, however, fail to provide support for either of these alternative explanations.

\medskip

 This paper contributes to an extensive body of literature investigating the relationship between ambiguity and the electoral behaviour of political parties. This literature can be categorized into two groups. First, there is a wide range of theoretical studies exploring why political parties may choose to be ambiguous in presenting their electoral programs to voters. These studies consider factors such as voters' risk aversion or the intensity of their preferences \citep{Shepsle1972, AlesinaCukierman1990, AragonesNeeman2000, AragonesPostlewaite2002, Laslier2006}, the importance of maintaining flexibility in office \citep{AragonesNeeman2000, Kartiketalt2017}, the significance of electoral competition in multi-party systems \citep{BrauningerGiger2018}, context-dependent voting patterns (Callander and Wilson, 2008), the influence of policy-motivated donors \citep{AlesinaHolden2008}, and whether political parties possess information about the median voter's position \citep{Glazer1990}. While the primary contribution of this paper is empirical, it connects with this theoretical literature in two ways. Firstly, it provides a theoretical explanation for the main empirical result by extending \cite{Glazer1990}'s model. Secondly, the paper explores whether certain theoretical models from this literature can account for the primary empirical result.
 
 In addition to the theoretical literature, there is a vast body of empirical research examining position blurring by political parties. For instance, \cite{Han2020} used data on party positions and public opinion on major political issues in Western Europe and found that, in polarized environments, political parties present clearer positions on the issues they primarily focus on, but less clear positions on secondary issues. \cite{BrauningerGiger2018} estimated ambiguity from electoral manifestos and found evidence that platforms become more ambiguous as the preferences of two key stakeholders, the voting public and the party's core constituency, diverge. Some studies have also utilized experimental data and found evidence that ambiguity can be a winning strategy \citep{TomzVanHouweling2009} and that it is more popular among non-centrist voters when one of the candidates is a known centrist \citep{TolvanenTremewanWagner2022}. While these papers offer valuable evidence regarding the factors influencing position blurring by political parties, they do not directly address how this blurring relates to the ideological position of parties on the political spectrum. Only a few papers, to our knowledge, \cite{Rovny2012, Rovny2013}, \cite{loprokschslapin2016}, and \cite{Praprotnik2017}, have explored this relationship directly, making them closely related to our study. 
 
 \cite{Rovny2012, Rovny2013} explores how political parties strategically employ ambiguity in elections where multiple issues are at play. \citeauthor{Rovny2012} argues that parties tend to emphasize dimensions where they hold extreme positions while blurring their stance on others. \citeauthor{Rovny2012} supports this theory with cross-sectional empirical evidence from over 100 political parties across 14 countries in 2006. In comparison to \citeauthor{Rovny2012}, although our paper also examines the relationship between ambiguity and centrism and finds an inverse relationship between the two, it differs on three important points: it expands the analysis to cover various time periods, more parties, and more countries; it proposes much more robust econometric evidence regarding the existence of this relationship, and it presents an alternative theory that does not focus on the notion of single-issue parties, which we argue is consistent with more robust and broader evidence.

\cite{Praprotnik2017} examines ambiguity in electoral competition in Austria and suggests that extremist parties, being niche parties, are penalized for presenting vague and ambiguous programs. In contrast, centrist parties, not being niche parties, have more incentives to be ambiguous. \citeauthor{Praprotnik2017} links this hypothesis to the government status of parties, arguing that government parties have incentives to decrease clarity in their campaign strategies compared to parties in opposition. Consequently, extremist parties in government might be equally or even more ambiguous than centrist parties in opposition. In comparison to \citeauthor{Praprotnik2017}, our paper also examines the relationship between ambiguity and centrism, but expands the analysis beyond Austrian political parties and niche extremist parties.

\cite{loprokschslapin2016} propose a method to estimate the ideological clarity of political parties based on party manifestos, and apply it to 74 parties in four countries. They find a positive correlation between their estimates of ideological clarity and the level of ideological extremism of political parties in those countries. They suggest that centrist parties may find ambiguous positions advantageous because they can appeal to a larger segment of the electorate located in the center of the political spectrum. 
This study expands upon  \citeauthor{loprokschslapin2016}'s research by including more countries and a larger sample of over 200 parties, resulting in more robust findings. Additionally, we explore alternative explanations and provide empirical evidence to support a new explanation that complements and extends the previous work.\footnote{It is worth noting that  \citeauthor{loprokschslapin2016} do not extensively delve into their primary explanation. Specifically, they observe that extreme voters tend to view ideological ambiguity as a sign of weakness or inadequate commitment to their cause, resulting in reduced support for extreme parties when they adopt ambiguous positions. The rationale behind why this behaviour appears to be specific to extreme voters warrants further investigation and remains an intriguing aspect to explore.}

The paper follows the following structure: Section \ref{sect_data_empirical} provides a comprehensive overview of the data utilized in the study and outlines our empirical approach. Section \ref{mainresults} presents the key findings of our analysis. In Section \ref{mechanisms1}, we put forth our primary explanation for the observed results. Section \ref{mechanisms2} then explores various alternative explanations. Lastly, Section \ref{Conclusion} concludes.


\section{Data and Empirical Strategy}
\label{sect_data_empirical}

\subsection{Data}
\label{sect_data}

This paper's analysis primarily relies on the Chapel Hill Expert Survey (CHES) dataset \citep{Jolly2022}, which provides party position data on ideology for numerous national parties across various European countries.\footnote{This data is publicly available and can be found at \url{https://www.chesdata.eu}}  The CHES dataset incorporates information from multiple surveys conducted in 1999, 2002, 2006, 2010, 2014, 2017 and 2019. These surveys involve a consistent assessment of party positions by a substantial number of experts, encompassing general left-right ideology, economic left-right orientation, and social values (GAL-TAN).\footnote{The CHES survey invites each expert to assess the position of each political party across three dimensions: (i) its overall ideological stance, (ii) its ideological stance on economic issues, involving classifications based on the party's position on matters such as privatization, taxes, regulation, government spending, and the welfare state, and (iii) its views on social and cultural values.}  The first set of variables employed in this study is derived from these assessments of party positions. In the more recent surveys (2017 and 2019), the CHES introduced a series of questions to gauge the extent of ambiguity in the establishment of these positions by political parties. The second set of variables used in this study is based on these measures of blurriness.\footnote{\label{footnoteSD}As a complement to our blurriness measure, we also consider the standard deviation of expert assessments. I thank a referee for suggesting this extension. While this alternative measure allows for a larger sample size, it is employed solely as a robustness check for our baseline results.  We see it as reflecting variance among expert assessments rather than directly measuring blurriness. Though significant differences in expert assessments plausibly indicates blurred ideology, there are instances where experts differ greatly without perceiving platforms as very blurred. This may explain the relatively low correlation between the two blurriness measures: 0.5 for the economic dimension and 0.14 for the social and cultural values dimension. Nonetheless, our baseline results remain robust with the use of this alternative measure.}

Given our focus on examining the relationship between ideology and blurriness, our analysis is restricted to the years 2017 and 2019. We merge the available data on party positions in the economic and social values dimensions with information on the associated degree of blurriness.  The resulting sample consists of 202 political parties, spanning across two time periods. This comprehensive dataset incorporates assessments from a minimum of 25 experts, covering at least two ideological dimensions.

All the data in this study is available at the expert-party-year level. In our primary analysis, we aggregate this information by averaging across experts for the same party-year. However, all our results remain robust when examined at the individual expert-party-year level.\footnote{Due to changes in the identity of experts across surveys, it is not feasible to construct a panel structure at the expert-party-year level. Consequently, in terms of identification, the econometric specification at the party-year level is equivalent to that at the expert-party-year level.}  The information on party positions is measured on a scale ranging from 0 to 10. In the economic dimension, a value of 0 represents an extreme left position, while a value of 10 indicates an extreme right position.\footnote{In the economic dimension, the CHES defines left and right as follows: parties on the economic left advocate for an active role of the government in the economy, while parties on the economic right support a reduced role for the government.} In the social values dimension, a value of 0 corresponds to a pro-libertarian/postmaterialist stance, whereas a value of 10 represents a pro-traditional/authoritarian perspective.\footnote{The CHES defines these terms as follows: while `Libertarian' or `postmaterialist' parties favor expanded personal freedoms, for example, abortion rights, divorce, and same-sex marriage,  `traditional' or `authoritarian' parties reject these ideas in favor of order, tradition, and stability, believing that the government should be a firm moral authority on social and cultural issues.}  Similarly, the data on blurriness is also measured on a scale ranging from 0 to 10. A value of 0 indicates no blurriness at all, whereas a value of 10 signifies extreme blurriness.\footnote{Specifically, the CHES asks each expert `how blurry was each party’s position on ...'.}

In addition to the main dataset, we incorporate supplementary databases to gain deeper insights into the underlying mechanisms that drive our primary findings. These additional databases comprise country-level data on economic outcomes, such as GDP per capita sourced from the International Monetary Fund (IMF), as well as party-level data on electoral outcomes, including the seat-share of parties in each election obtained from the ParlGov project \citep{DoringHuberManow2022}. A more detailed description of these databases will be presented in subsequent sections of the study.

\subsection{Empirical Strategy}\label{sec_empiricalstrategy}

In our initial econometric specification, we model the outcome $blurriness_{ipct}$, which represents the average expert opinion on the level of blurriness in the political position of party $p$ regarding issue $i$ in country $c$ during year (or wave) $t$, as
\begin{equation}
\label{eqbaseline}
blurriness_{ipct}=\beta_0+\beta_1 \times position_{ipct} +\beta_2 \times position^2_{ipct}+\gamma_{ct}+\epsilon_{ipct}
\end{equation}
where $position_{ipct}$ is the average opinion of experts on the position of party $p$  regarding issue $i$ in country $c$ during period $t$. $position^2_{ipct}$ refers to the square of this position. The term  $\gamma_{ct}$ are country $\times$ year fixed effects, and $\epsilon_{ipct}$ is the error term.\footnote{We also estimate and report results from a model using data at the expert level. However, as mentioned earlier, since the experts change with each survey, including expert fixed effects is not possible. Consequently, for identification purposes, this model is equivalent to the one specified in Equation (\ref{eqbaseline}).}  As we will explain later, the inclusion of $\gamma_{ct}$ is crucial in our identification strategy, as it allows us to control for various country-related factors that may change over time, such as population, quality of national institutions, characteristics of party systems, and political polarization.  We also introduce political party fixed effects in certain specifications to account for party-related factors that remain constant over time. Including these fixed effects greatly strengthens the robustness of our results, indicating that the relationship exists not only within countries but also within parties.\footnote{It is worth noting that the inclusion of these fixed effects significantly reduces the sample size, which is why we do not use this specification as our main model. However, all our results remain robust even when these fixed effects are included.}

 The coefficients of interest are $\beta_1$ and $\beta_2$,  which capture the relationship between the parties' positions on each issue and the perceived blurriness of these positions by experts. 
 
 \medskip

To facilitate the analysis of mechanisms, we also estimate different versions of the following equation:
\begin{equation}
\label{eqalt}
blurriness_{ipct}=\alpha_0+\alpha_1 \times centrism_{ipct}+\gamma_{ct}+\eta_{ipct}
\end{equation}
where  $centrism_{ipct}$ represents the difference between 5 and the level of extremism of party $p$ on issue $i$ in country $c$ during period $t$. Extremism is measured as the absolute value of the difference between 5 and the average expert opinion on each party's position. The coefficient of interest, $\alpha_1$, captures the relationship between the level of ideological centrism of party positions on each issue and how experts perceive the blurriness of these positions.\footnote{An alternative way to define centrism is by considering the median of the position distribution within each dimension for every party system. I thank a referee for suggesting this alternative. In the following section we show that estimates using either measure are identical.}

\medskip

This paper is descriptive in nature, and achieving a perfect identification strategy is challenging given the research question's nature. Specifically, the models presented in Equations (\ref{eqbaseline}) and (\ref{eqalt}) may not establish causal effects due to at least two endogeneity concerns. 

Firstly, there could be omitted factors driving the association between blurriness and party position. While we control for any time-varying variables at the country level and include party fixed effects, certain time-varying variables at the party level might still influence the results. For instance, the level of internal dissent within each party could be one such variable: centrist parties may experience more internal dissent, and internal dissent may lead to more blurred positions. Additionally, the age of a party might matter, with older parties possibly adopting more centrist positions and having less blurred stances. Government status is another potential factor: centrist parties might be more frequently in government, and governing parties may be rewarded for presenting clear ideological stances. In the next section, we show that the main results remain robust even after controlling these potential time-varying party-level factors, which helps alleviate the first concern.

Secondly, a more critical issue is the potential presence of simultaneity bias in the models of Equations (\ref{eqbaseline}) and (\ref{eqalt}). This bias could occur if the perception of a party's position as blurred (by experts) increases the likelihood of identifying that position as centrist. To address this concern, we adopt an instrumental variables (IV) approach by using lagged values of position (or centrism) as instruments. The effectiveness of this estimation strategy  hinges on two primary assumptions. Firstly, the relevance assumption, which we verify with our data. Secondly, and more importantly, the assumption that conditional on fixed effects and controls, lagged values of position (or centrism) affects blurriness only through their contemporaneous values (i.e. the exclusion restriction assumption). Although our IV strategy does not resolve all endogeneity concerns, we think it provides a reliable approach as long as the lagged values of position do not influence blurriness through unaccounted time-varying party-level factors.  In addition to providing supplementary evidence that aligns with this assumption, it is important to highlight that we could not identify any such factors that could undermine the validity of this IV strategy (given the substantial number of controls and fixed effects included).


\section{Main Results}
\label{mainresults}

\begin{figure}[H]
     \centering
             \caption{Party position on economic and social issues and blurriness}
        \label{fig1_blurrposition}
     \begin{subfigure}[b]{0.51\textwidth}
         \centering
         \includegraphics[width=\textwidth]{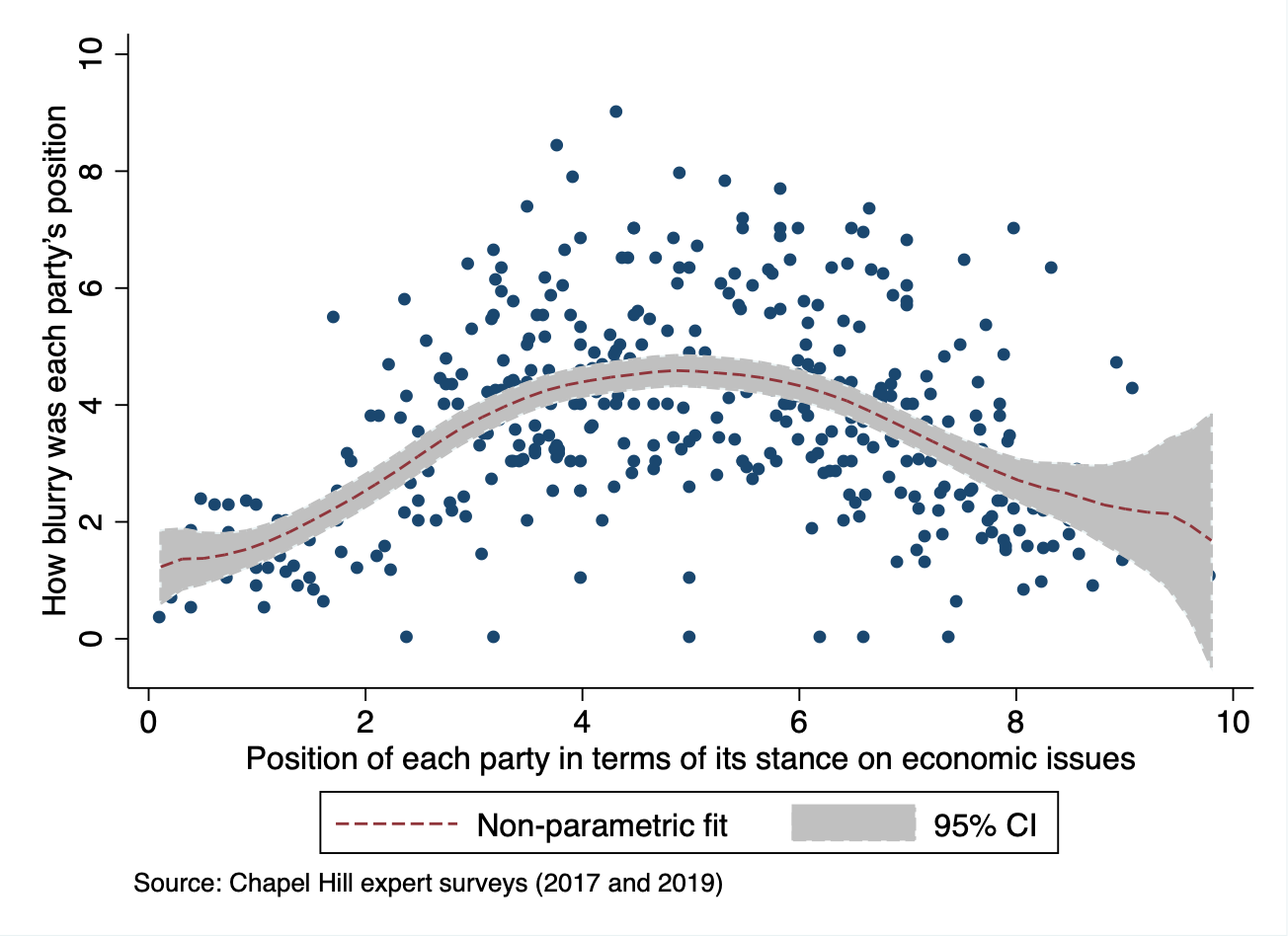}
         \caption{Economic issues}
         \label{fig1_blurrposition_a}
     \end{subfigure}
     \hspace{-1cm}.
     \begin{subfigure}[b]{0.51\textwidth}
         \centering
         \includegraphics[width=\textwidth]{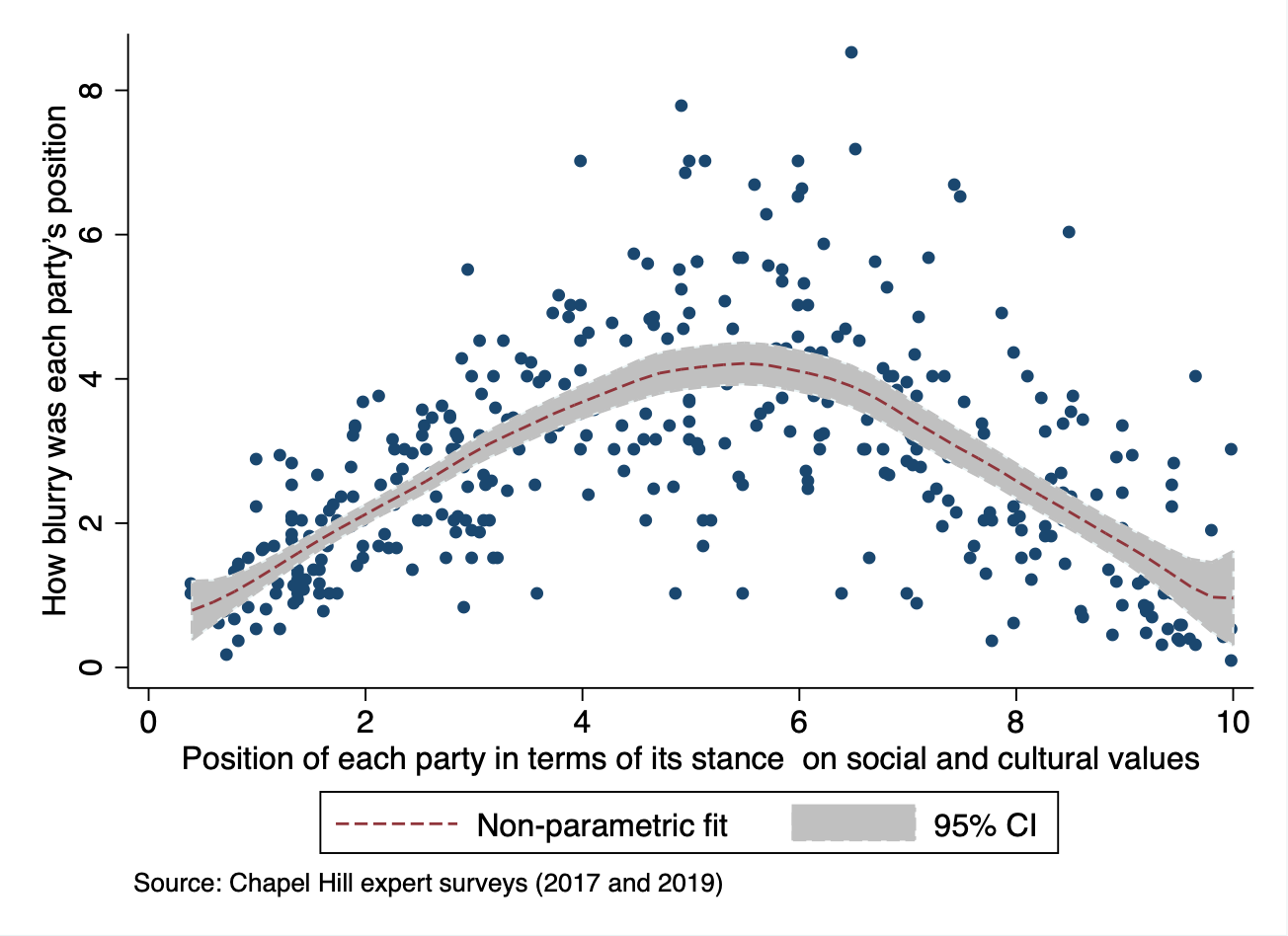}
         \caption{Social values}
         \label{fig1_blurrposition_b}
     \end{subfigure}
\end{figure}

\begin{figure}[H]
     \centering
             \caption{Centrism on economic and social issues and blurriness}
        \label{fig2_blurrcentrism}
     \begin{subfigure}[b]{0.51\textwidth}
         \centering
         \includegraphics[width=\textwidth]{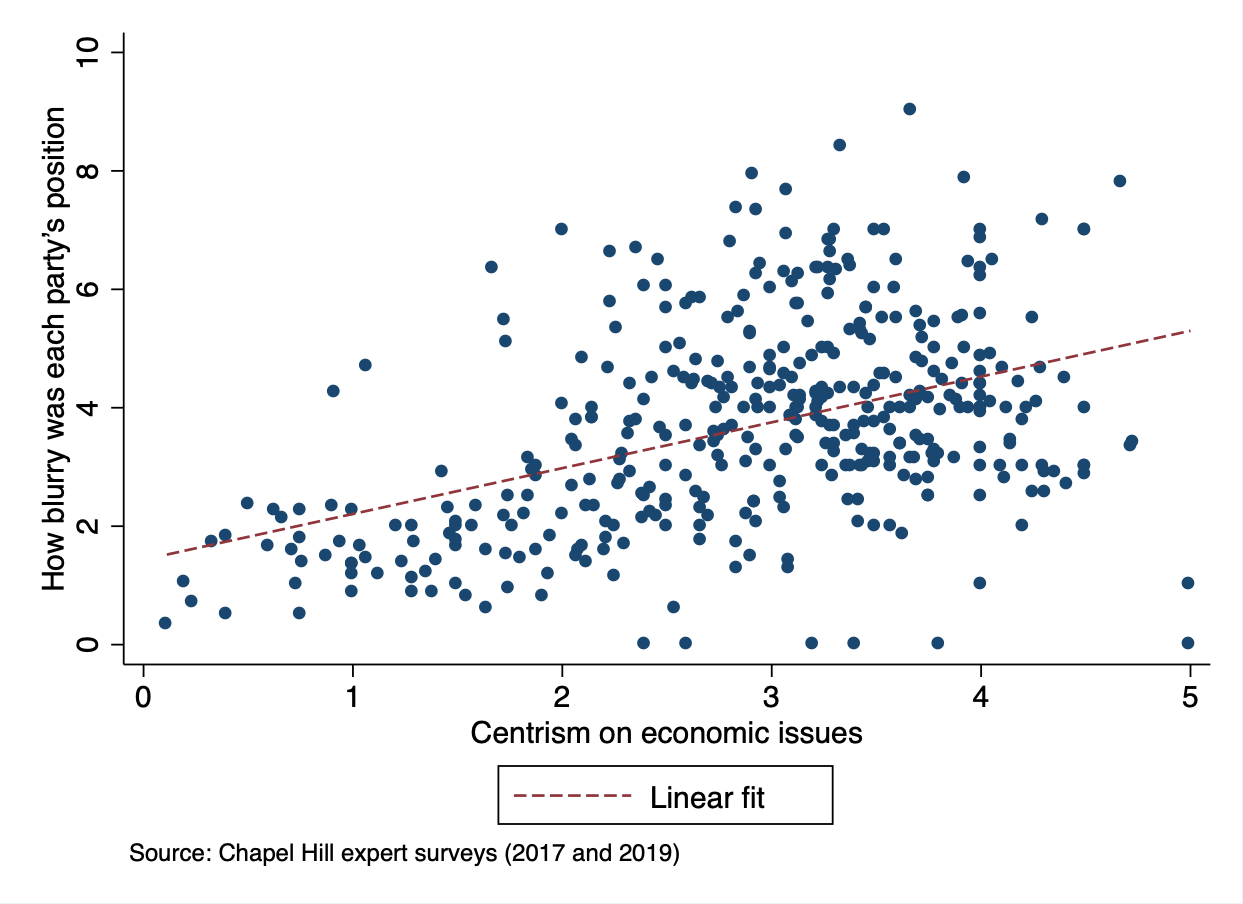}
         \caption{Economic issues}
         \label{fig2_blurrcentrism_a}
     \end{subfigure}
     \hspace{-1cm}.
     \begin{subfigure}[b]{0.51\textwidth}
         \centering
         \includegraphics[width=\textwidth]{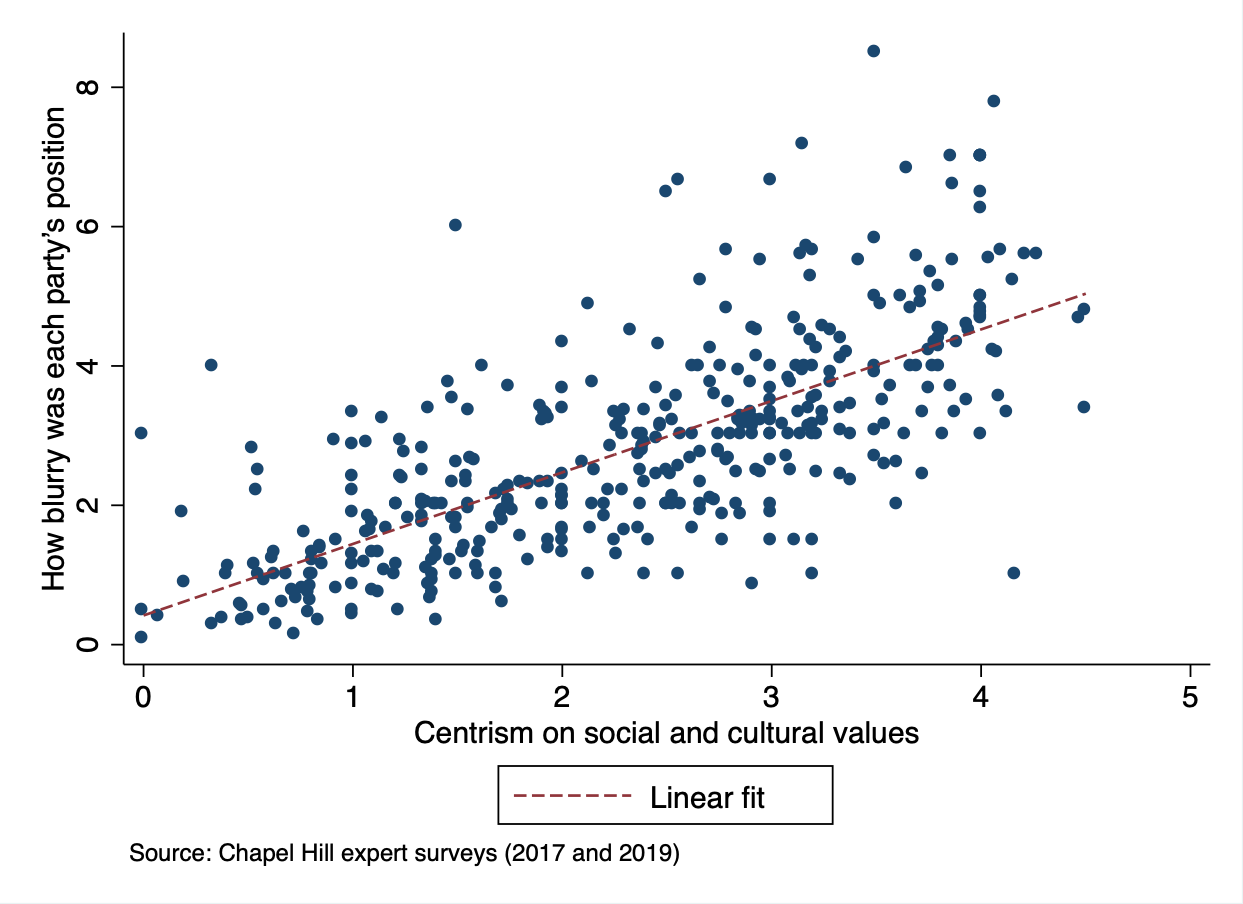}
         \caption{Social issues}
         \label{fig2_blurrcentrism_b}
     \end{subfigure}
\end{figure}

Figure \ref{fig1_blurrposition} presents the relationship between experts' opinions on each party's position on economic issues (Fig. \ref{fig1_blurrposition_a}) and social values (Fig. \ref{fig1_blurrposition_b}) and the perceived blurriness of these positions. Both figures exhibit a clear non-monotonic inverted U-shaped pattern. Notably, the peak of these curves occurs around the midpoint of the ideological position distribution (approximately 5), which corresponds to centrist political parties. Complementing these results, Figures \ref{fig2_blurrcentrism_a} and \ref{fig2_blurrcentrism_b} show a strong positive association between the level of centrism of each party's position (as defined in the previous section) and the level of blurriness of these positions, consistent with the non-monotonic inverted U-shaped relations found in Figures \ref{fig1_blurrposition_a} and \ref{fig1_blurrposition_b}.

Columns (1) to (6) in Table \ref{tab1_baseline} provide the estimates of Equations \ref{eqbaseline} and \ref{eqalt} for the outcomes analyzed in Figures \ref{fig1_blurrposition} (Panel A) and \ref{eqalt} (Panel B). Columns (1) to (4) report the estimates of Equation \ref{eqbaseline}, showing positive and negative coefficients for $\beta_{1}$ and $\beta_{2}$, respectively. All coefficients are individually and jointly statistically significant at conventional levels. Additionally, the peak points of the curves fall between 4.1 and 5.3, well within the range of ideological positions (from 1 to 10). These results confirm the inverted U-shaped relationships observed in Figures \ref{fig1_blurrposition_a} and \ref{fig1_blurrposition_b}. Columns (5) and (6) present the estimates of Equation (\ref{eqalt}), showing positive and statistically significant effects, which is consistent with the strong positive correlations seen in Figures \ref{fig2_blurrcentrism_a} and \ref{fig2_blurrcentrism_b}. Finally, columns (7) and (8) in Table \ref{tab1_baseline} report fixed effects OLS estimates of a hypothetical monotonic relationship between a party's position and blurriness. However, all these estimates are statistically insignificant at conventional levels, providing evidence for the existence of a non-monotonic relation.\footnote{Tables \ref{tableA1} and \ref{tab_centrismalt} in Appendix \ref{tables_A} present estimates of Equations (\ref{eqbaseline}) and (\ref{eqalt}) 
 using expert-level data and an alternative centrism measure based on the median position distribution (instead of the midpoint of the scale).  As expected, these estimates are virtually the same as those in Table \ref{tab1_baseline}.}

As a complement to the previous findings, Figure \ref{fig_blurrpositionSD} and Table \ref{tab_baselineSD} in Appendix \ref{tables_A} show estimates of Equation (\ref{eqbaseline}) for an alternative outcome: the standard deviation in expert assessments relative to party positions on the economic dimension (Figure \ref{fig_blurrpositionSD_a} and columns (1) to (4) of Table \ref{tab_baselineSD}) and the social values dimension (Figure \ref{fig_blurrpositionSD_b} and columns (5) to (8) of Table \ref{tab_baselineSD}). As discussed earlier (see footnote \ref{footnoteSD}), while we do not view this alternative measure of blurriness as a substitute for the baseline results, it serves as an important robustness check. Notably, the results in Figure \ref{fig_blurrpositionSD} and Table \ref{tab_baselineSD} align with our baseline findings, confirming the previously observed inverted U-shaped relationship.

\medskip

As discussed in the previous section, the estimates in Table \ref{tab1_baseline} may suffer from serious endogeneity bias. To address this concern, I re-estimate the models in Equations (\ref{eqbaseline}) and (\ref{eqalt}), while (i) controlling for potential confounding time-varying variables at the party level, and (ii) instrumenting  position and centrism with lagged values. 

Table \ref{tableA2} in Appendix \ref{tables_A} presents estimates of the baseline models while incorporating additional controls, namely, the level of internal dissent within each political party, the age of each party, and whether the party is in government. The results shown in Table \ref{tableA2} are statistically indistinguishable from those displayed in Table \ref{tab1_baseline}. As previously explained, this supplementary evidence reduces the likelihood of omitted variables bias in our baseline results. Table \ref{tableA3} in Appendix \ref{tables_A} provides instrumental variable (IV) estimates for the models described in Equations (\ref{eqbaseline}) and (\ref{eqalt}), employing lagged position (and centrism) as instruments. The table presents the IV estimates for the second stage, along with the F-statistic pertaining to the first stage. In addition to providing evidence consistent with the relevance assumption (e.g. showing a F-statistic consistently exceeding 10), Table \ref{tableA3} also shows estimates that are statistically equivalent to those presented in Table \ref{tab1_baseline}. As previously argued, this supplementary evidence reduces the likelihood of simultaneous bias in our baseline results and offers additional evidence against the existence of omitted variables bias.\footnote{As discussed in Section \ref{sec_empiricalstrategy}, a pivotal assumption in our instrumental variable (IV) specification is the exclusion restriction assumption. This assumption entails that, given fixed effects and controls, the influence of lagged centrism on blurriness operates solely through their contemporaneous values. As argued in Section \ref{sec_empiricalstrategy},  given the extensive number of fixed effects and controls included, we find it unlikely that a mechanism exists through which a party's centrism influences its blurriness level via a channel not captured by such fixed effects and controls. However, to increase the plausibility of this assumption, Table \ref{tableA3aa} in Appendix \ref{tables_A} offers fixed effects OLS estimations based on Equation (\ref{eqalt}), but incorporating lagged centrism. Notably, in alignment with the exclusion restriction assumption, our findings show that lagged centrism exhibits no substantial correlation with blurriness when controlling for contemporaneous centrism (see columns (3), (4), (7), and (8) of Table \ref{tableA3aa}).} 

While we acknowledge that the findings in Tables \ref{tableA2}, \ref{tableA3} and \ref{tableA3aa} do not entirely eliminate bias in the main results of this section, nor do they establish causality, at the very least, they indicate that any potential bias is likely to be small.


\section{Explanation: Uncertainty of Extremes}\label{mechanisms1}

So far we have documented a non-monotonic relationship in the form of an inverted U-shaped curve between the ideology of the parties and the degree of ideological blurriness or ambiguity. Moreover, through the use of various econometric models, we have presented arguments suggesting that the ideological position not only correlates with blurriness but likely exerts an influence on it. This result is important in and of itself. However, this result can be explained in several ways. In this section, we propose an explanation that we consider the most plausible and empirically examine its validity. While the analysis is exploratory in nature, it holds the potential to offer valuable insights.

Our proposed explanation is based on the idea that centrist political parties are perceived as a safer choice for voters compared to extremist parties. This perception stems from the tendency of extremist parties to propose untested and unconventional policies, while centrist parties tend to advocate for more established and moderate policies. Given the general risk aversion of the median voter, centrist parties inherently possess an advantage over extremists. As a result, centrist parties strategically exploit this advantage by adopting more ambiguous positions than extremist parties. This strategic approach enables them to align more closely with the ideal point of the median voter (a point often unknown by political parties), thereby increasing the probability of winning elections while incurring relatively lower costs.\footnote{Note that if ambiguity facilitates political parties in aligning their positions with the median voter, this explanation shares similarities with a finding in \cite{AragonesNeeman2000} (see Theorem 1.(ii) and 1.(iii)). We thank a referee for suggesting this connection.  \cite{AragonesNeeman2000} establishes that, in equilibrium, only extremist parties, competing against each other, can afford to adopt ambiguous platforms. As we will elaborate below, our model diverges from that proposed by \cite{AragonesNeeman2000} as it incorporates an initial asymmetry between extremist and centrist parties.}

Before introducing the formal model that serves as the basis for this explanation, it is worthwhile to delve into its implications and assumptions. Firstly, the explanation is highly general and stylized, aiming to capture a widespread phenomenon observed in a relatively large sample. In this sense, we acknowledge the possibility of alternative explanations for more specific contexts.\footnote{In connection with this, it is worth noting that this explanation aligns with the hypothesis that centrist political parties, exhibiting ambiguity, can be seen as catch-all parties. Thus, an interpretation of our main results based on centrists predominantly being catch-all parties is consistent with our explanation.}

Secondly, regarding the assumptions, it's noteworthy that some are consistent with existing empirical evidence. Research on the party-voter linkage indicates that voters are often uninformed about political parties' issue positions \citep{AdamsEzrowSomerTopcu2011, AdamsEzrowSomerTopcu2014}. Additionally, voters tend to dislike ambiguity from parties \citep{Danielle2019}. There is also evidence that the median voter’s position is usually not well-known to parties \citep{AbouchadStoetzer2020, LindvallRuedaZhaiFORTH}.

Thirdly, regarding the assumption that all voters are risk-averse, while evidence shows variations in voters' attitudes towards certain relevant dimensions \cite[][]{EhrlichMaestas2010, KamSimas2010, Kam2012, EcklesKametalt2014},  to our knowledge, there is no evidence suggesting differences fundamentally linked to voters' ideological orientation. Importantly, relevant evidence suggests that voting for change or populist options may depend on disparities in voters' traits related to their risk attitudes \cite[][]{SteenbergenSiczek2017, Morisi2018}.\footnote{This aspect, solely rooted in the demand-side and not in equilibrium, remains unexplored in our model and presents an interesting potential extension.}

Finally, concerning the assumption that political parties intentionally employ ambiguity as an electoral strategy, there is indirect yet substantial evidence backing this claim. For example, \cite{SomerTopcu2015} show empirically that broad appeal strategies help parties win votes by convincing voters they align closely with their preferences. Moreover, additional empirical evidence employing diverse methodologies support this premise \cite[][]{BrauningerGiger2018, Rovny2012, loprokschslapin2016}.\footnote{While most of the existing literature supports the idea that ambiguity is deliberate and strategic, some non-strategic sources for ambiguity have also been proposed. An example is the existence of intra-party divides \citep{LehrerLin2020}, which we believe does not affect our results as they are robust to controlling for this factor.}

\subsection{Model}\label{secc_model}

We now present a straightforward formalization of our previous intuition. The model we propose extends  \cite{Glazer1990}'s model to scenarios where uncertainty associated with ambiguous political party behaviour is asymmetric.

\subsubsection{Description of the game} 

In this model, we consider two political parties, competing to maximize their probability of winning. One party is centrist ($C$), and the other is extremist ($E$). Each party can propose a policy $s_{J}$ to the voters, where $J\in\{C,E\}$ and $s_{J} \in \mathbb{R}$. Both parties are uncertain about the median voter's ideal point, but they know its probability distribution. Similarly, voters are uncertain about the actual policy that will be implemented by the winning party, knowing only the probability distribution for this policy. For simplicity, we assume these distributions are uniform. Parties also have the option to be ambiguous, where not being ambiguous means proposing a policy $s_{J}\in S_{J}$, and being ambiguous means refraining from selecting any value from a potentially different set $S^{a}_{J}$, where $S_{J}\subseteq S^{a}_{J}$ and $\#S_{J}\leq \#S^{a}_{J}<\infty$ (with $\# S$ denoting the cardinality of $S$).

Crucially, we assume that $\#S^{a}_{C}<\#S^{a}_{E}$, indicating that if both types of parties are ambiguous, voters are ex ante more uncertain about the policy implemented by the winning party if it is the extremist party. Additionally, we assume that $S_{C}=S^{a}_{C}$, which is consistent with our main assumption, and that  $S_{C}=S_{E}$, meaning that if both parties decide not to be ambiguous, the uncertainty experienced by voters is ex ante the same.  To simplify the calculations, we define $S_{C}=S_{E}=S^{a}_{C}=\{-k,-1,0,1,k\}$ and $S^{a}_{E}=\{-l,-k,-1,0,1,k,l\}$, where $1<k<l$. Finally, we assume that the outcome of the election is determined by the preferences of a risk-averse median voter, whose preferences can be represented by the utility function $u(x)=-x^{2}$.\footnote{This model can be extended in several dimensions, such as the spaces of possible alternatives or the preferences of the median voter. However, for simplicity, and given the primary empirical focus of this paper, we present the simplest version of the model.} 

\subsubsection{Equilibrium} 

In Appendix \ref{model_A}, we show that in equilibrium, when the centrist party is unable to adopt highly extreme policies, the centrist party chooses ambiguity, while the extremist party chooses not to be ambiguous. Specifically, we establish that  (i) if $k^{2}<\frac{3}{2}$, the centrist party opts to be ambiguous while the extremist party decides not to be ambiguous, and (ii) if $k^{2}>\frac{3}{2}$, both parties choose not to be ambiguous.

The  main idea of the proof and the underlying rationale for this result arises from  the strategic decisions of parties to either adopt an ambiguous stance or not. To delve into these scenarios, it is crucial to assess the median voter's expected utility in each situation, considering that both parties share the primary objective of maximizing their probability of victory, and the median voter theorem applies.  In the first scenario, both parties choose ambiguity. Here, the median voter's expected utility is $u(p_{C}|a)=-\frac{2(k^{2}+1)}{5}$ when a centrist party wins, and $u(p_{E}|a)=-\frac{2(l^{2}+k^{2}+1)}{7}$ when an extremist party wins. Importantly, $u(p_{C}|a)>u(p_{E}|a)$,  indicating the median voter's preference for the centrist party when  both parties are ambiguous. A second scenario arises when exactly one party specifies a position. If a party $J$ declares a position, given that $S_{J}=\{-k,-1,0,1,k\}$, with a probability of $\frac{3}{5}$, this party selects a position within one unit of the median voter's ideal point, ensuring the median voter a payoff of at least $-1$. A third scenario involves both parties specifying a position. In this case, since $S_{C}=S_{E}$ (implying the median voter's indifference), each party has an equal probability of winning.
 
Building upon this analysis, Appendix \ref{model_A} explores how equilibrium actions depend on the parameter $k$ in each of the aforementioned scenarios. We find that for sufficiently small values of $k$—when the policy alternatives available to the centrist party are not excessively radical—we expect centrist parties to exhibit more ambiguity than extremist parties. This expectation arises because, for sufficiently small values of $k$, the extremist party can only defeat the centrist party by specifying a position, given the higher uncertainty faced by the median voter if the extremist party wins. In contrast, the centrist party, benefiting from lower uncertainty associated with its potential victory, faces less pressure to specify a position and, by being ambiguous, aligns itself closer to the median voter's ideal point.

Our model, along with the explanation of the results in Section \ref{mainresults} that we have proposed, centers on how voters perceive extremist political parties. Specifically, it hinges on the uncertainty that voters face when considering the consequences of choosing extremist parties compared to centrist parties. Since extremist parties often propose untested and unconventional policies, while centrist parties present more established and moderate policies, voters perceive centrist political parties as less risky than extremist parties. We argue that due to the inherent asymmetry concerning this uncertainty and the fact that the median voter is generally risk-averse, centrist parties have more incentives than extremist parties to strategically use ambiguity to win elections.

  \subsection{Evidence supporting the uncertainty of extremes explanation}
  
We now assess the empirical plausibility of the explanation presented in the preceding subsection. Our analysis focuses on instances where extremist political parties are plausibly perceived to entail greater risks when compared to centrist parties. Our objective is to investigate whether, in accordance with the proposed mechanism, centrist parties also exhibit a heightened level of ambiguity. We center our analysis on countries marked by substantial fluctuations in recent economic growth, particularly those where an extremist party has held power in the recent past. While in such scenarios it may be expected that all parties are, on average, less ambiguous, we hypothesize that in such circumstances, voters anticipate lower economic risks under the governance of a centrist party. Consequently, this anticipation induces extremist parties to adopt less ambiguous stances than the centrist parties.

To empirically test this hypothesis, we present estimates derived from various specifications detailed in Table \ref{tab4_uncertainty}. Columns (1) to (3) examine economic issues, while columns (4) to (6) focus on social values. In all the specifications, we introduce an interaction term between each party's level of economic or social centrism and the lagged GDP growth variance in each country.\footnote{We employ data on real GDP growth from the IMF, publicly available at \url{https://www.imf.org/external/datamapper/NGDP_RPCH@WEO/OEMDC/ADVEC/WEOWORLD}.}  Additionally, in columns (2), (3), (5), and (6), we further interact this specification with a dummy variable indicating parties not in the government.\footnote{To construct this variable, we use party-level data on electoral outcomes from the ParlGov project, publicly available at \url{https://www.parlgov.org/data-info/}. We use the fact of not obtaining the largest number of seats in national parliaments as a proxy for parties not being part of the government cabinet. We assume that if a party does not secure the most seats, it is less likely to join a government coalition and be viewed as responsible for government actions. Although this measure is clearly noisy (as there are several cases where parties with the most votes do not govern), we believe the bias created by this noise is downward.} 

The result in column (1) shows an estimated coefficient of the first interaction that is positive but statistically insignificant. Column (2) indicates that for the economic dimension, in countries with larger GDP growth variation in the last year, this effect becomes more pronounced and statistically different from zero when centrist parties were not part of the government in the last year. Column (3) demonstrates that these results remain robust and have larger magnitudes when including party fixed effects. Finally, columns (4) to (6), which focus on social values, do not show estimated coefficients of the interactions that are statistically significant. This is not surprising, given the emphasis on scenarios primarily related to economic aspects, such as countries experiencing high variation in economic growth.\footnote{In Table \ref{tab_uncertaintyGDPalt} in Appendix \ref{tables_A}, we explore the robustness of the results in Table \ref{tab4_uncertainty} using an alternative measure of GDP growth variance, a dummy variable equal to one if the GDP growth variance of each country in each year is greater than the median of the distribution of GDP growth variances of all countries in that same year. We thank a referee for suggesting this alternative measure. Table \ref{tab_uncertaintyGDPalt} shows statistically identical results to those in Table \ref{tab4_uncertainty}.}

The findings in Table \ref{tab4_uncertainty} support the `uncertainty of extremes' hypothesis we have put forth. Specifically, they suggest that the increased ambiguity of centrist parties may be linked to the perception that extreme parties pose higher risks for voters. While these results, along with those of the previous subsections, strengthen our preference for the proposed hypothesis, they do not entirely rule out alternative explanations that could also contribute to the results of Section \ref{mainresults}. Nevertheless, the results in Table \ref{tab4_uncertainty} indicate that the hypothesis presented in this section could be a valuable component of a comprehensive explanation for the findings obtained in Section \ref{mainresults}.

\section{Other Possible Explanations}\label{mechanisms2}

\subsection{Single-issue politics}\label{mechanisms21}

An alternative explanation for our findings is rooted in the idea that in multidimensional political competitions, certain parties strategically adopt blurred and centrist positions on specific issues in order to appeal to a broader voter base. Simultaneously, these parties may prioritize other, potentially single, issues.  If these `single-issue' parties prioritize issues outside of the economic or social dimensions, it is likely that they would display decreased ambiguity and less centrism on those particular issues.\footnote{This electoral strategy has been referred to as a 'broad-appeal strategy' by \cite{SomerTopcu2015}, and there is ample evidence of its implementation by several political parties in Europe. Notable studies examining this phenomenon include \cite{Rovny2012, Rovny2013}, \cite{Han2020}, and \cite{RovnyPolk2020}.}

To assess the plausibility of this explanation, we examine one of its immediate implications: centrist parties that are identified as blurred in the economic and social dimensions should not exhibit centrist positions in the dimension(s) they prioritize.\footnote{
As discussed in Section \ref{sect_data}, we have data on blurriness only for the economic and social dimensions. Therefore, if single issue parties prioritize a dimension  beyond these two, we cannot observe the level of ambiguity of these parties regarding that specific dimension.This is clearly a limitation of our analysis.  However, as we will elaborate later, we do possess information about the degree of centrism exhibited by these parties across a wide range of dimensions. Thus, our analysis stands valid on the assumption that these parties will exhibit less centrism in their prioritized dimensions.} Although empirically assessing this hypothesis is challenging due to the difficulty in identifying single-issue political parties, we can utilize information from several additional and significant policy dimensions to explore whether the findings from the previous section can be explained by the presence of single-issue political parties that focus on these dimensions. By including a diverse set of dimensions in the analysis,\footnote{These dimensions encompass a wide range of issues, such as immigration policy, multiculturalism policy, economic redistribution, environmental sustainability, spending vs. taxes, deregulation of markets, intervention in the economy, civil liberties vs. law and order, social lifestyle, religious principles in politics, ethnic minority rights, nationalism, urban/rural interests, protectionism, decentralization, anti-Islam rhetoric, anti-elite rhetoric, and European integration.} it is unlikely that, on average, single-issue political parties would concentrate on a dimension that is not correlated with any of them.

\begin{figure}[H]
     \centering
             \caption{Centrism across policy dimensions: economic issues vs social values}
        \label{fig3_econvsgaltan}
     \begin{subfigure}[b]{0.51\textwidth}
         \centering
         \includegraphics[width=\textwidth]{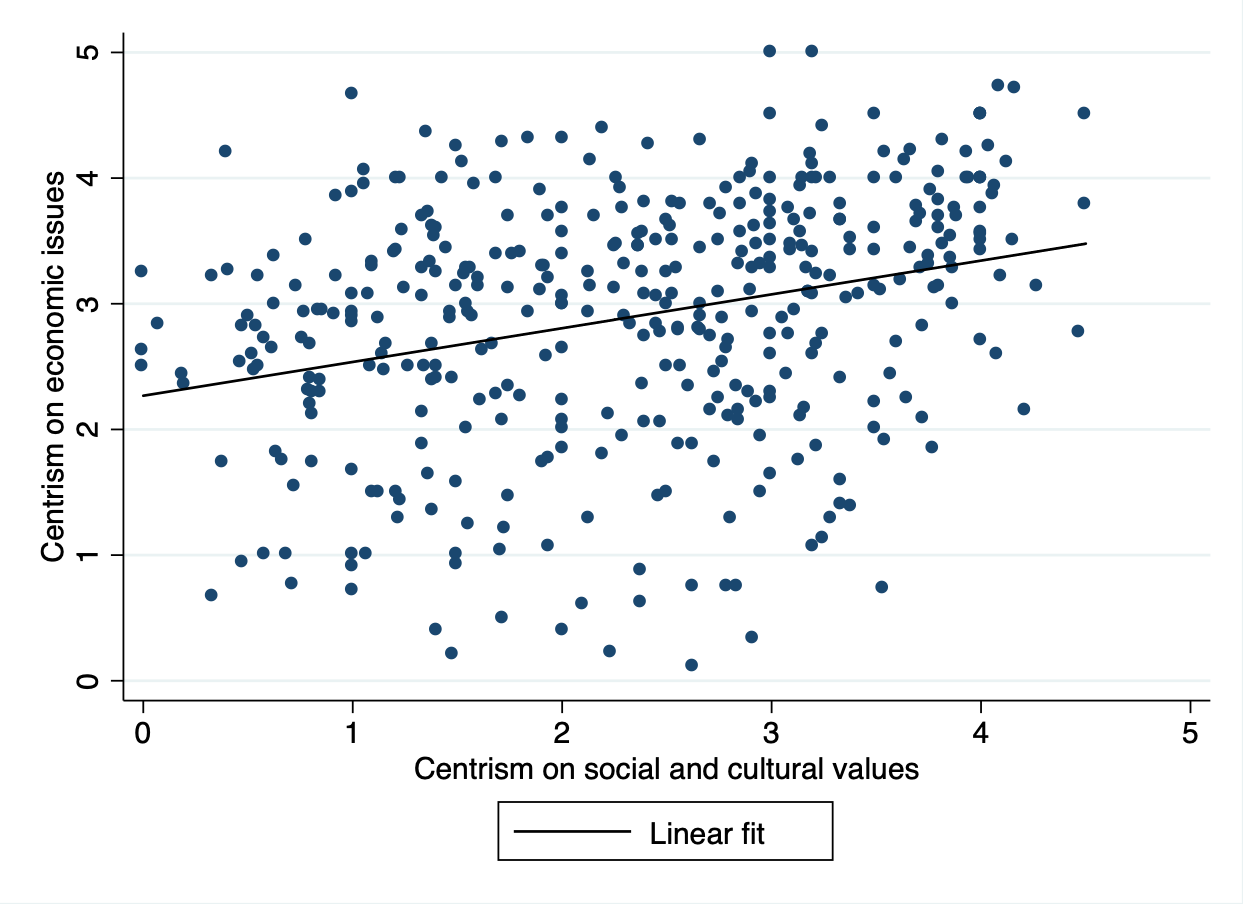}
         \caption{Simple correlation}
         \label{fig3_econvsgaltan_a}
     \end{subfigure}
     \hspace{-1cm}.
     \begin{subfigure}[b]{0.51\textwidth}
         \centering
         \includegraphics[width=\textwidth]{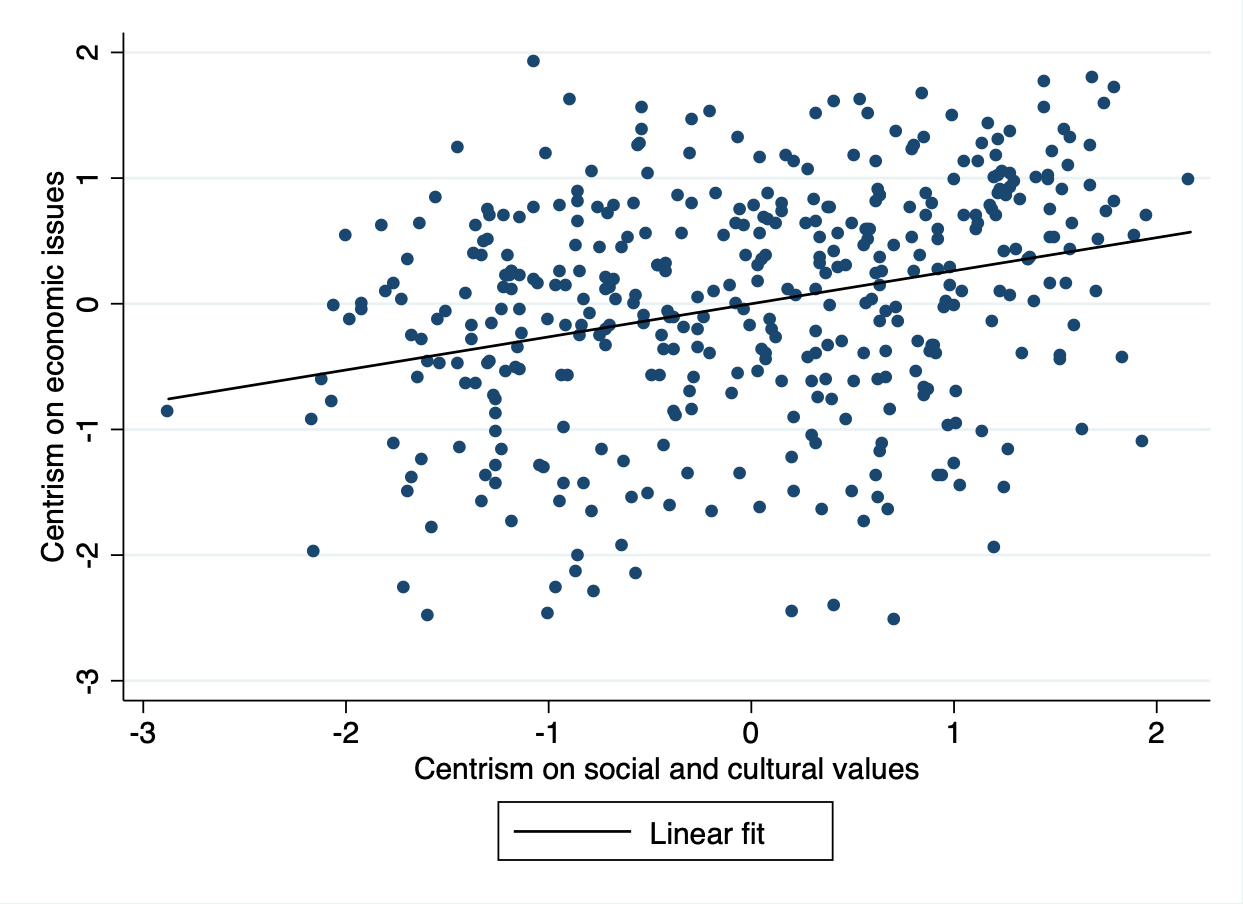}
         \caption{Partial correlation}
         \label{fig3_econvsgaltan_b}
     \end{subfigure}
\end{figure}

\begin{figure}[H]
     \centering
             \caption{Centrism across policy dimensions: economic issues vs other issues}
        \label{fig4_econvsothers}
     \begin{subfigure}[b]{0.48\textwidth}
         \centering
         \includegraphics[width=\textwidth]{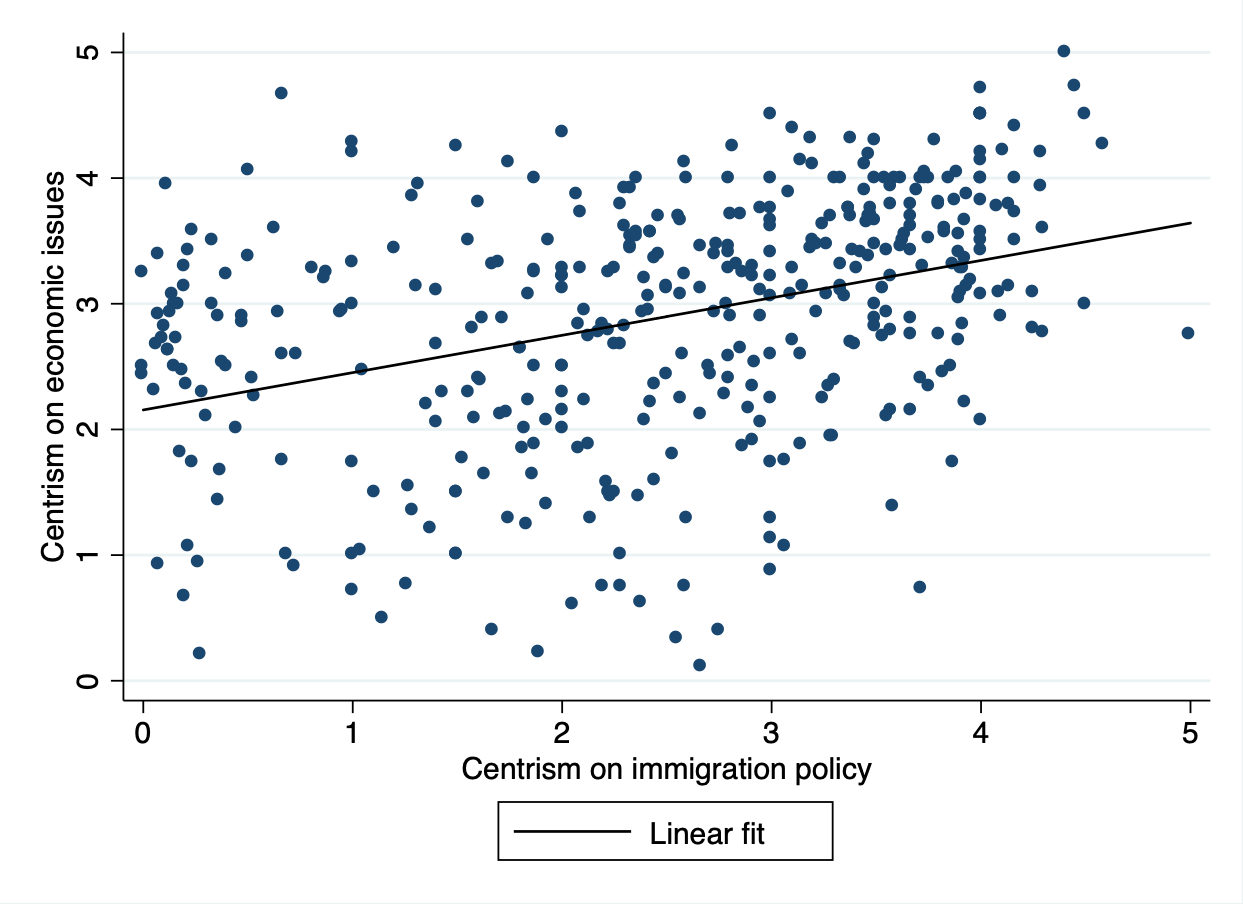}
         \caption{Immigration policy}
         \label{fig4_econvsothers_a}
     \end{subfigure}
     \begin{subfigure}[b]{0.48\textwidth}
         \centering
         \includegraphics[width=\textwidth]{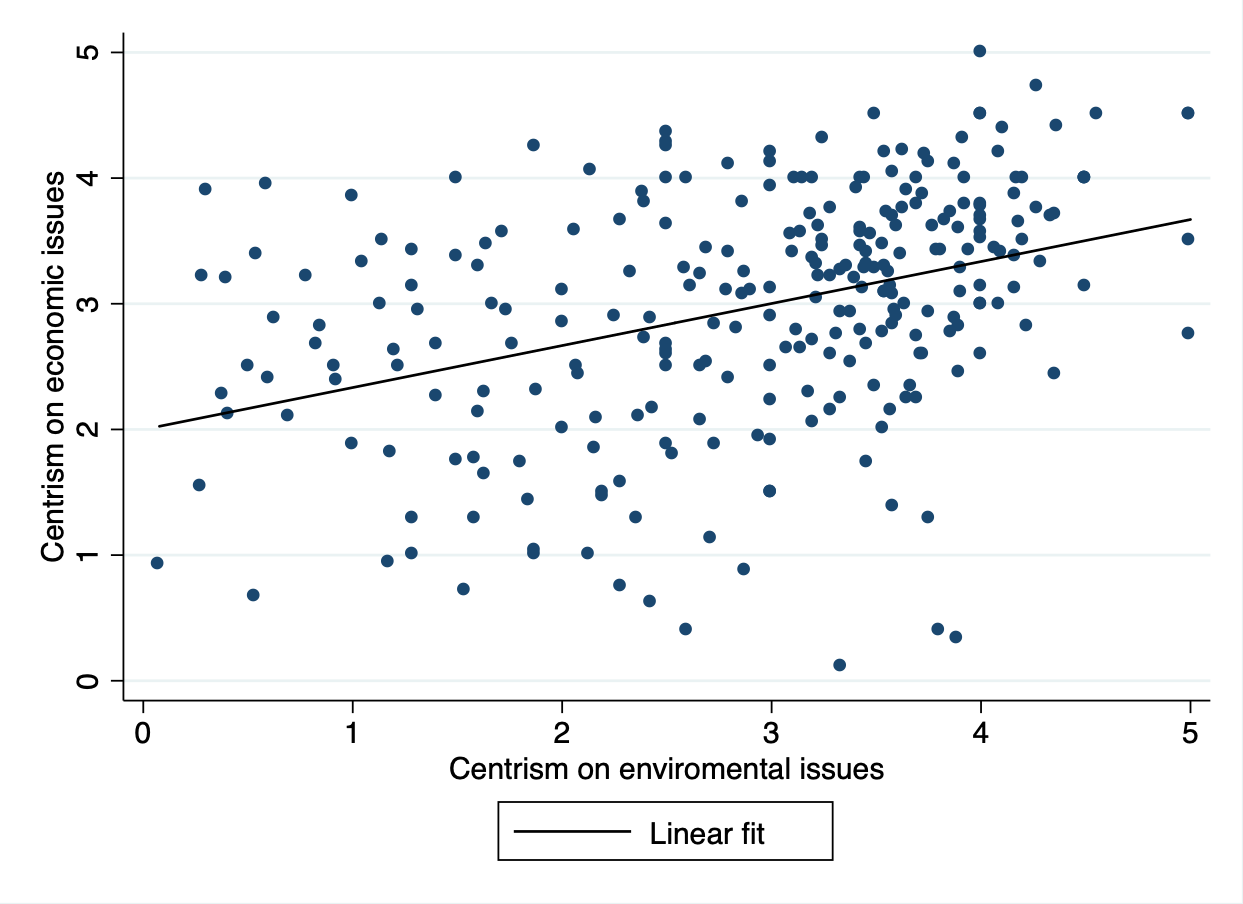}
         \caption{Enviromental issues}
         \label{fig4_econvsothers_b}
     \end{subfigure} 
     
     \vspace{0.1cm}
     
          \begin{subfigure}[b]{0.48\textwidth}
         \includegraphics[width=\textwidth]{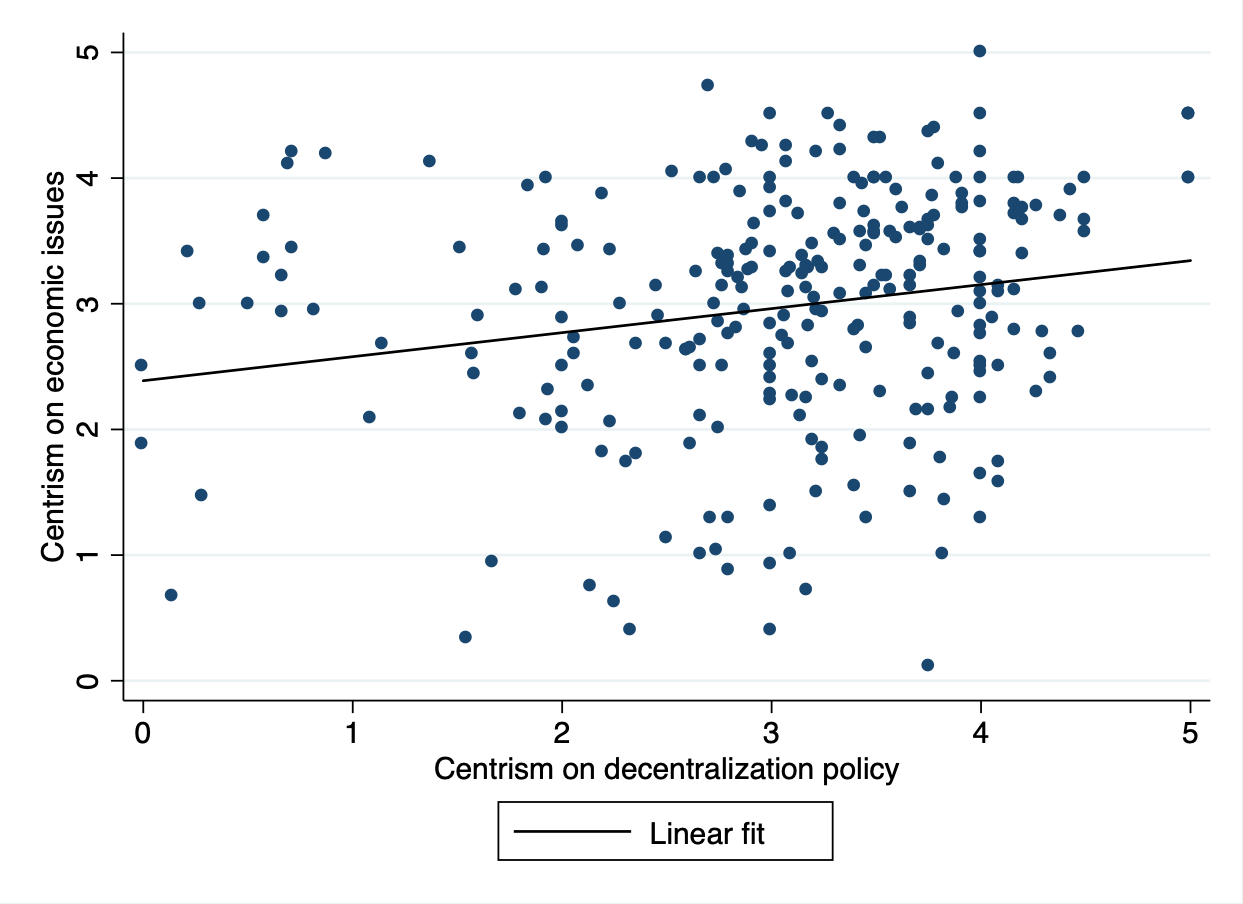}
         \caption{Decentralization policy}
         \label{fig4_econvsothers_c}
     \end{subfigure}
     \begin{subfigure}[b]{0.48\textwidth}
         \centering
         \includegraphics[width=\textwidth]{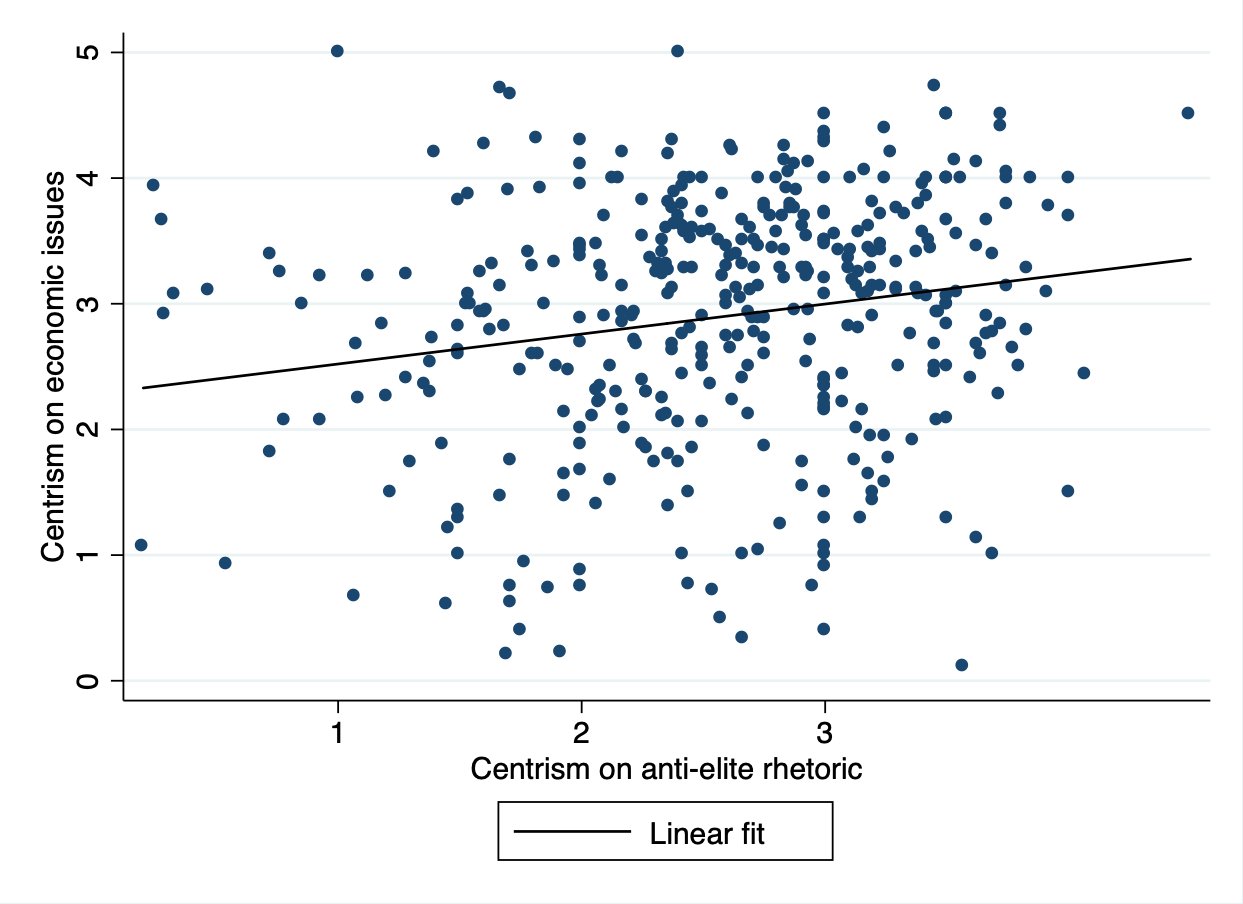}
         \caption{Anti-elite rhetoric}
         \label{fig4_econvsothers_d}
     \end{subfigure}
\end{figure}

\begin{figure}[H]
     \centering
             \caption{Centrism across policy dimensions: social values vs other issues}
        \label{fig5_galtvsothers}
     \begin{subfigure}[b]{0.48\textwidth}
         \centering
         \includegraphics[width=\textwidth]{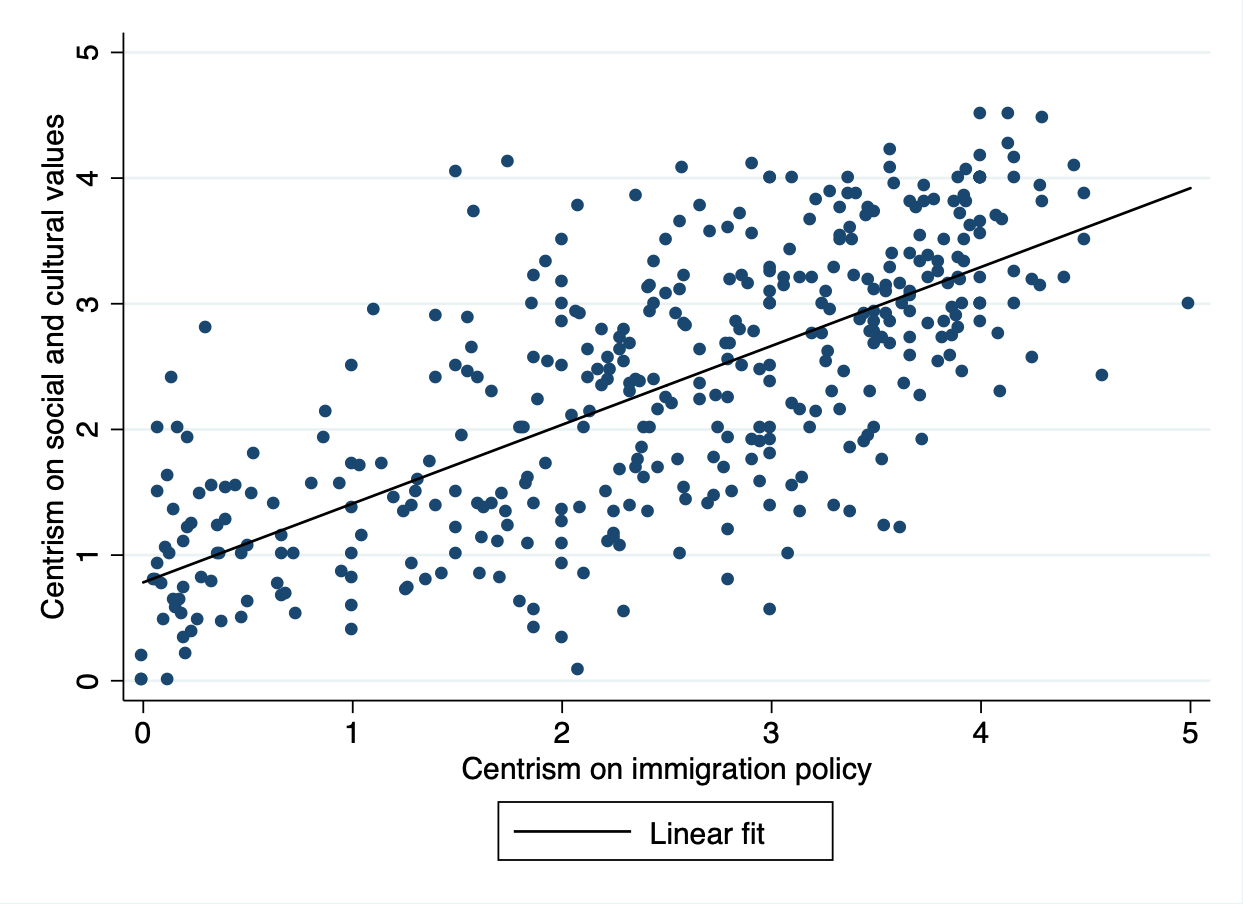}
         \caption{Immigration policy}
         \label{fig5_galtvsothers_a}
     \end{subfigure}
     \begin{subfigure}[b]{0.48\textwidth}
         \centering
         \includegraphics[width=\textwidth]{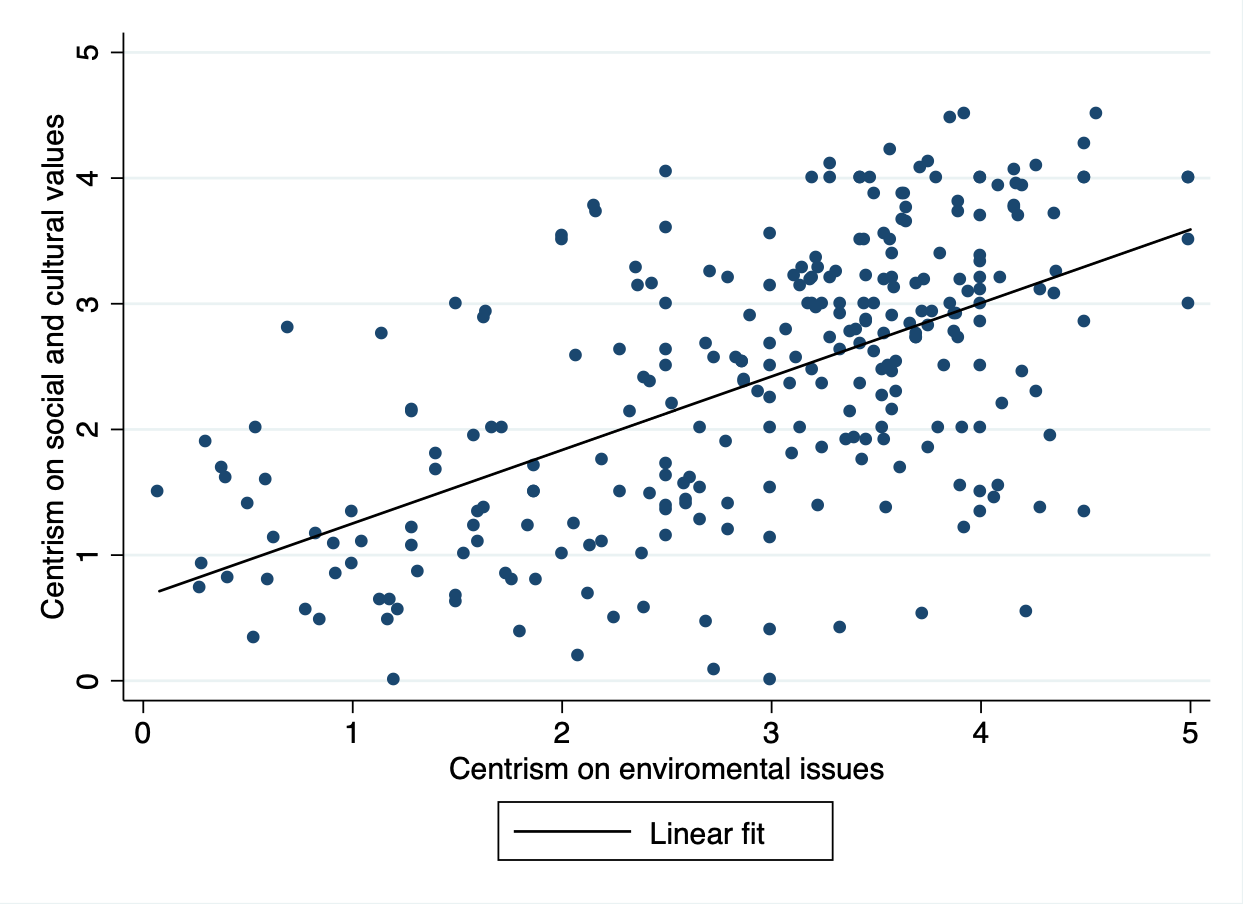}
         \caption{Enviromental issues}
         \label{fig5_galtvsothers_b}
     \end{subfigure} 
     
     \vspace{0.1cm}
     
          \begin{subfigure}[b]{0.48\textwidth}
         \includegraphics[width=\textwidth]{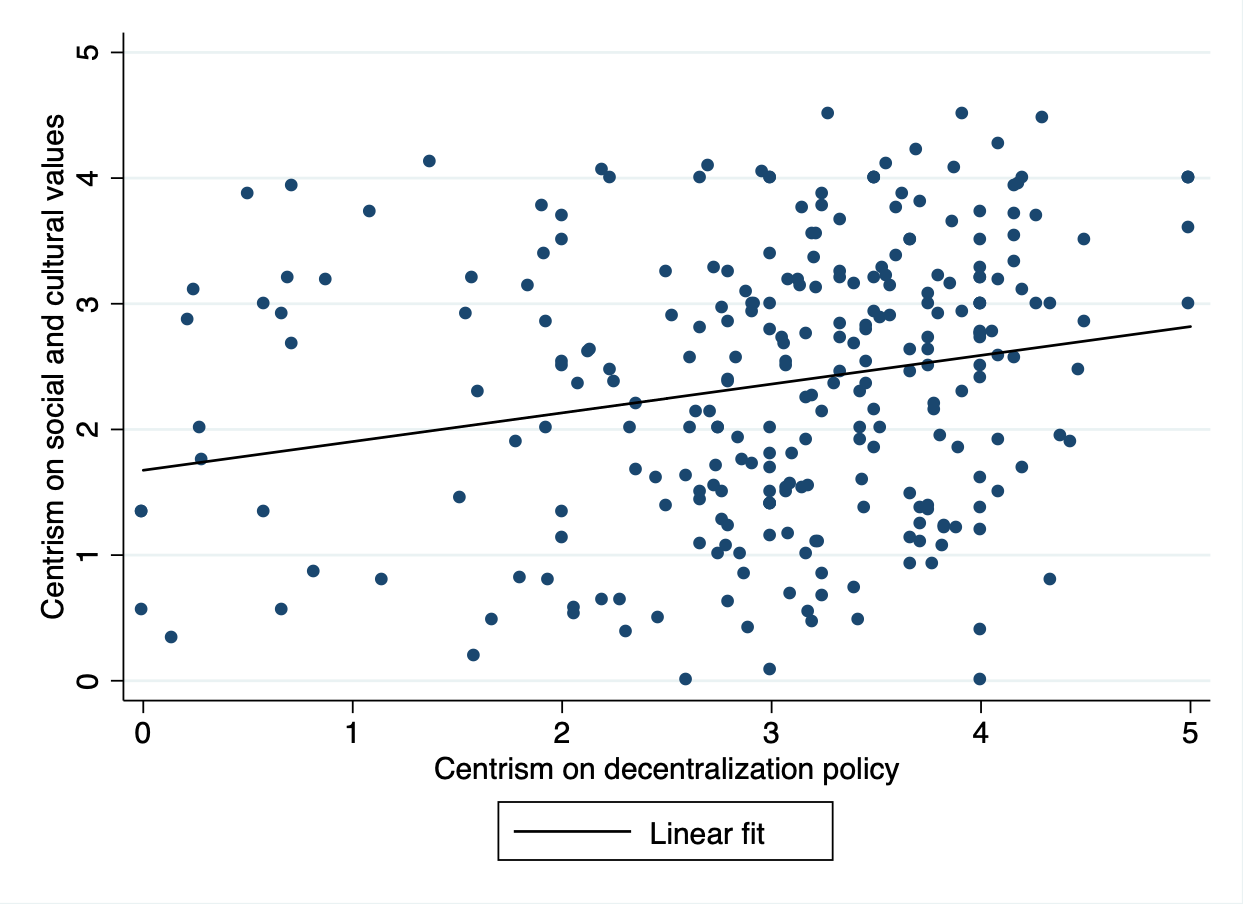}
         \caption{Decentralization policy}
         \label{fig5_galtvsothers_c}
     \end{subfigure}
     \begin{subfigure}[b]{0.48\textwidth}
         \centering
         \includegraphics[width=\textwidth]{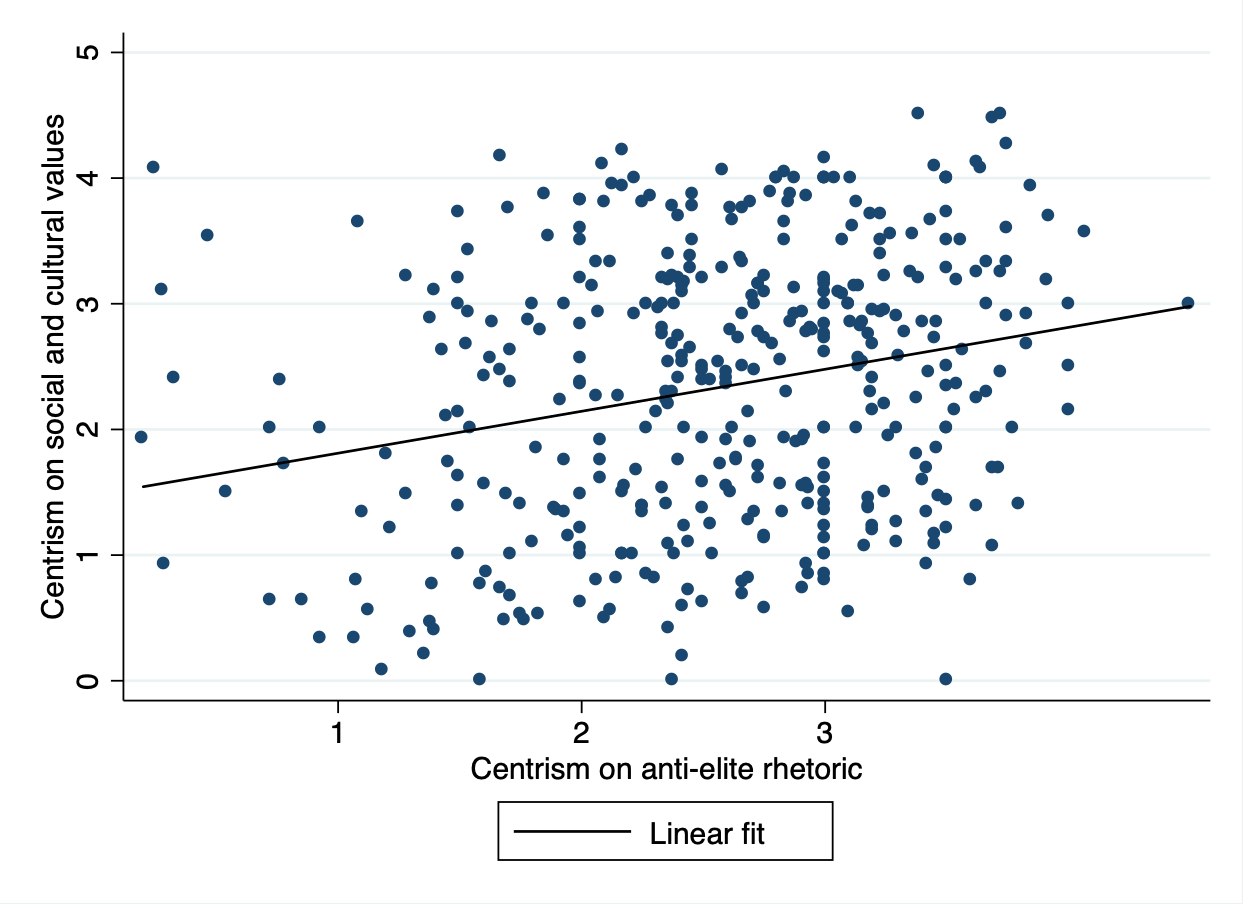}
         \caption{Anti-elite rhetoric}
         \label{fig5_galtvsothers_d}
     \end{subfigure}
\end{figure}

Figures \ref{fig3_econvsgaltan} to \ref{fig5_galtvsothers} illustrate the correlation between the level of centrism of political parties in the economic and social dimensions, as well as their correlation with four other ideological dimensions not directly related to economics or social values. Figure \ref{fig3_econvsgaltan} specifically focuses on the relationship between the centrism of parties in the economic and social dimensions, demonstrating a robust positive correlation. Similarly, Figures \ref{fig4_econvsothers} and \ref{fig5_galtvsothers} depict the correlation between the centrism of parties in the economic and social dimensions with four additional dimensions: immigration policy, environmental sustainability, decentralization policy, and anti-elite rhetoric. Importantly, all the figures demonstrate significant positive correlations.

To further support these findings, Table \ref{tableA5} in Appendix \ref{tables_A} expands upon the analysis presented in Figures \ref{fig3_econvsgaltan} to \ref{fig5_galtvsothers}, encompassing 18 additional policy dimensions and reporting the statistical significance of the correlations. The table reveals two key findings. Firstly, the correlations observed in Figures \ref{fig3_econvsgaltan} to \ref{fig5_galtvsothers} are statistically significant. Secondly, and notably, the pattern described in these figures extends to 15 other policy dimensions.\footnote{These other policy dimensions are: multiculturalism policy, economic redistribution, spending vs taxes, deregulation of markets, intervention in the economy, civil liberties vs. law and order, social lifestyle, religious principles in politics, ethnic minority rights, nationalism, and protectionism.}

\begin{figure}[H]
     \centering
             \caption{Centrism across policy dimensions: role of European integration, urban/rural interests, anti-Islam rhetoric}
        \label{fig6_econgaltanvseuint}
     \begin{subfigure}[b]{0.48\textwidth}
         \includegraphics[width=\textwidth]{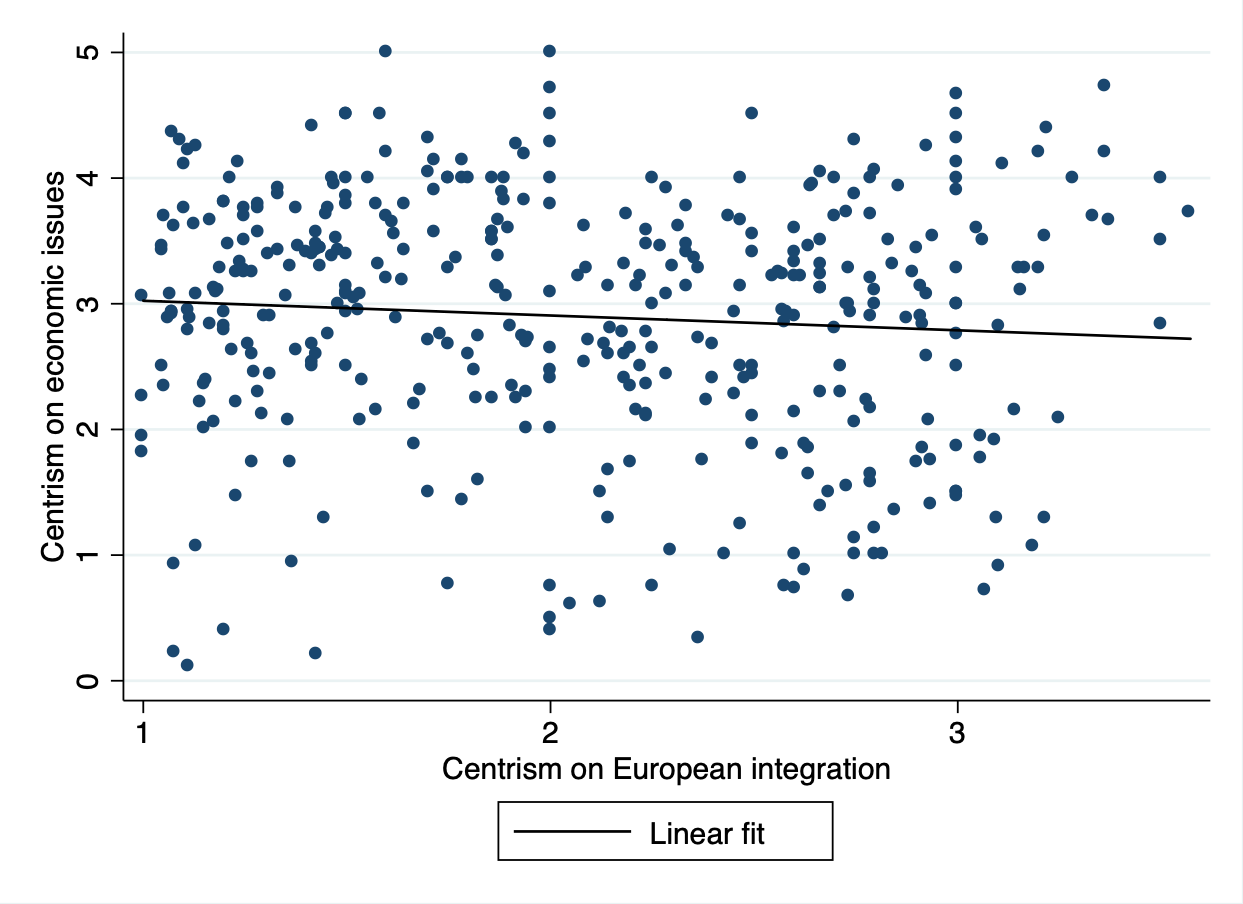}
         \caption{Economic issues vs European integration}
         \label{fig6_econgaltanvseuint_a}
     \end{subfigure}
     \begin{subfigure}[b]{0.48\textwidth}
         \includegraphics[width=\textwidth]{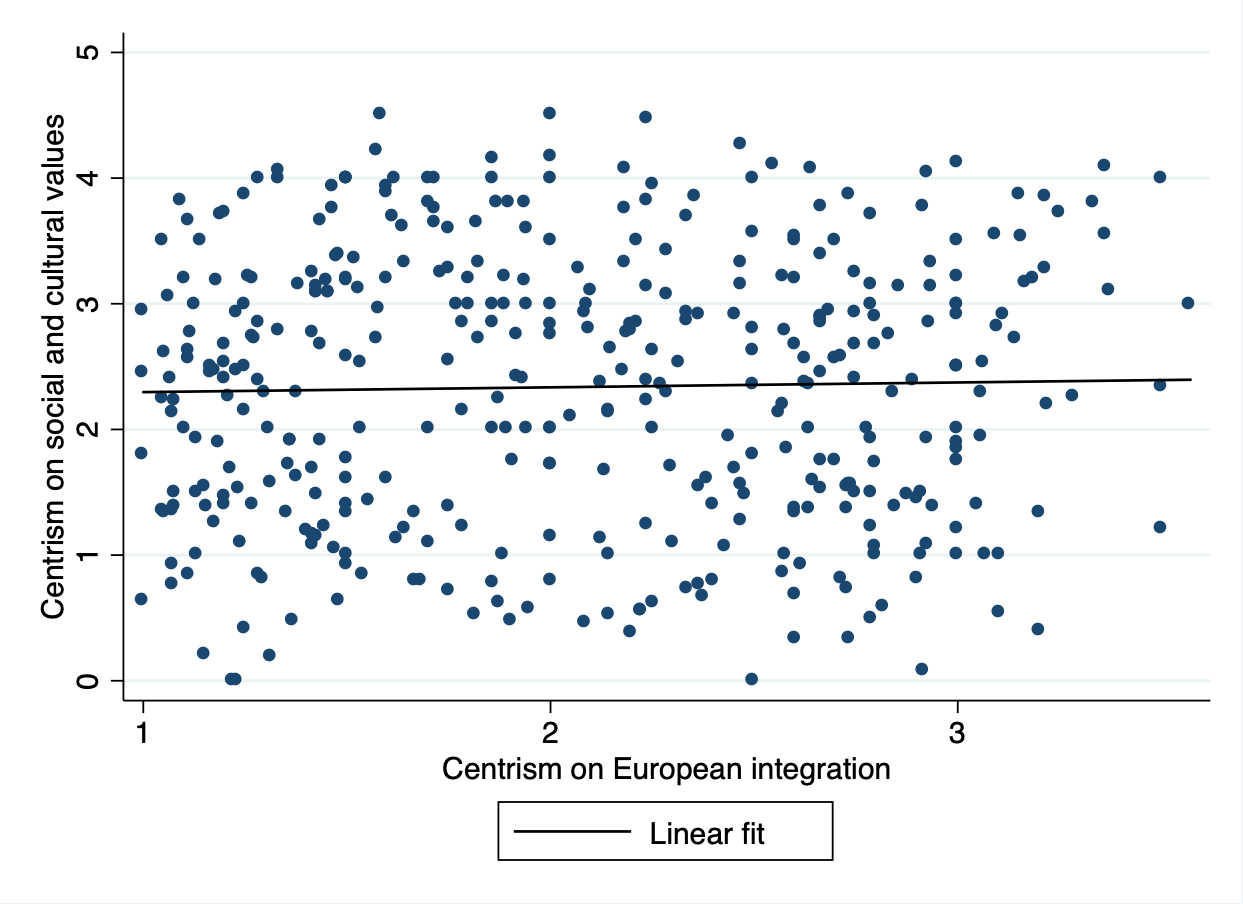}
         \caption{Social values vs European integration}
         \label{fig6_econgaltanvseuint_b}
     \end{subfigure}
     
          \vspace{0.1cm}
          
          \begin{subfigure}[b]{0.48\textwidth}
         \includegraphics[width=\textwidth]{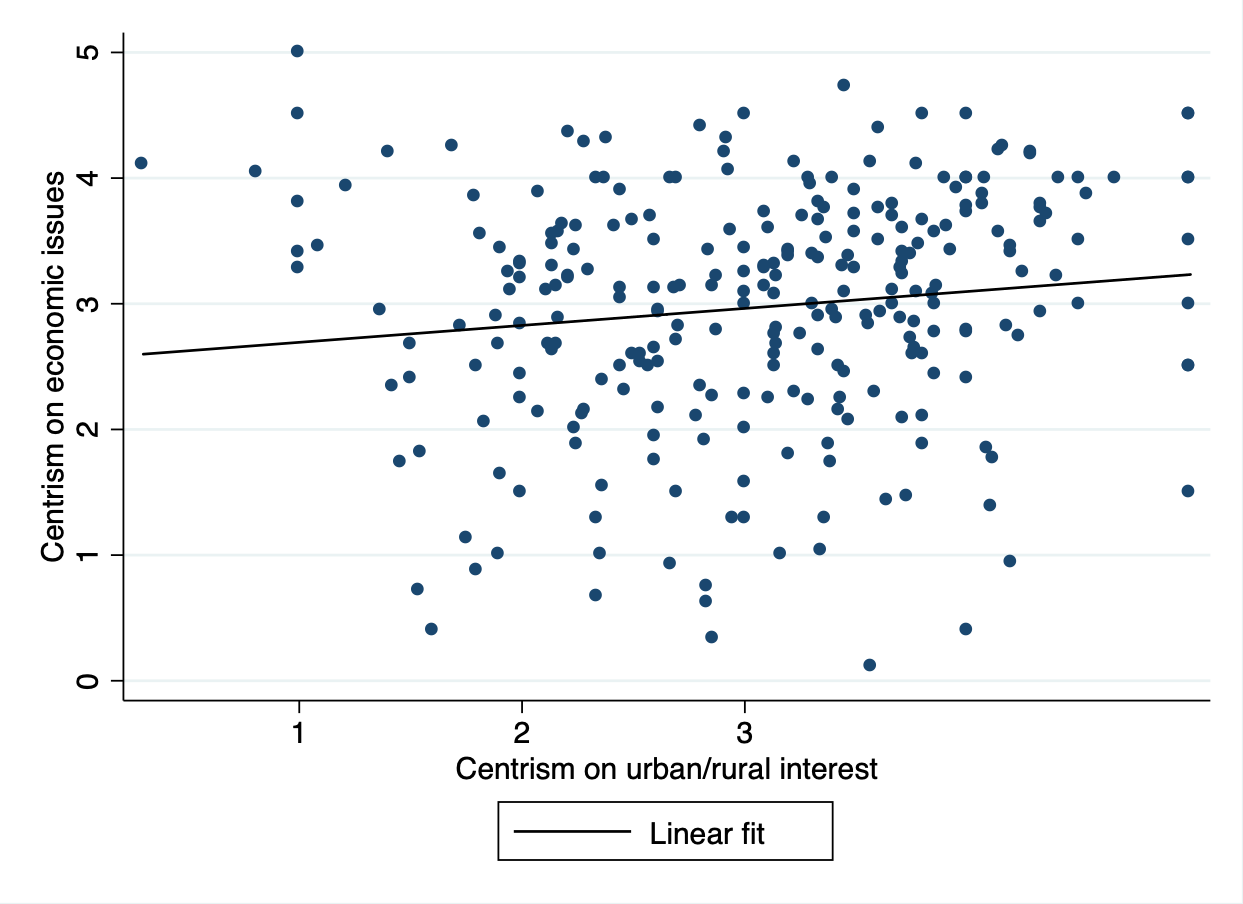}
         \caption{Economic issues vs urban/rural interests}
         \label{fig6_econgaltanvseuint_c}
     \end{subfigure}
          \begin{subfigure}[b]{0.48\textwidth}
         \includegraphics[width=\textwidth]{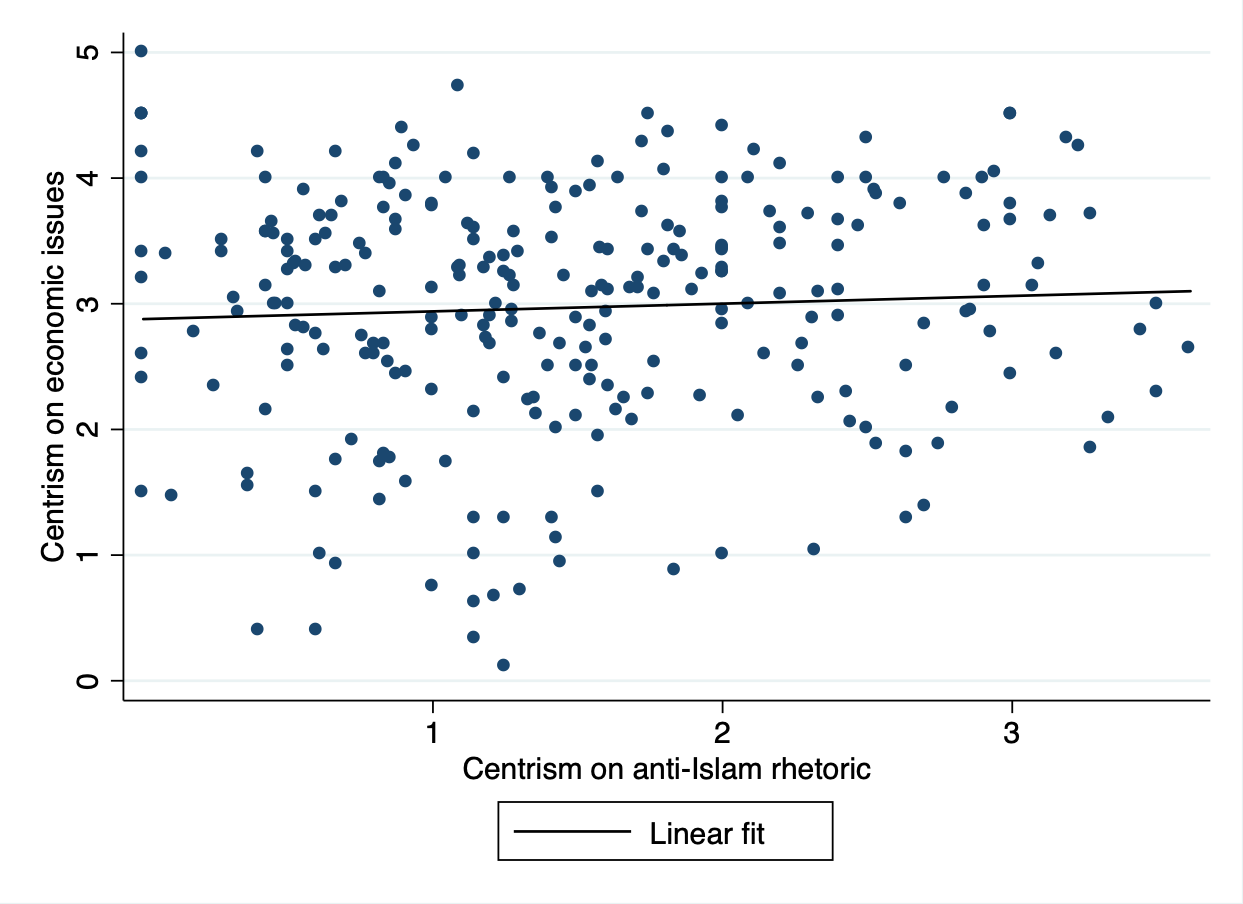}
         \caption{Economic issues vs anti-Islam rhetoric}
         \label{fig6_econgaltanvseuint_d}
     \end{subfigure}    
     
\end{figure}

The results presented in Figures \ref{fig3_econvsgaltan} to \ref{fig5_galtvsothers} and Table \ref{tableA5} challenge the single-issue hypothesis, providing evidence that centrist parties in both economic and social dimensions exhibit centrist ideological positions across 15 policy dimensions. This suggests that maintaining a centrist and blurred stance is not indicative of prioritizing other issues. However, it is important to acknowledge that these results are not consistent across all dimensions for which data is available. Specifically, Table \ref{tableA5} reveals two noteworthy observations. Firstly, the degree of centrism displayed by political parties in the economic dimension does not correlate with their level of centrism in the dimensions of European integration, urban/rural interests, and anti-Islam rhetoric. Secondly, the level of centrism demonstrated by parties in the social dimension does not align with their level of centrism in the dimension of European integration.\footnote{Figure \ref{fig6_econgaltanvseuint} plots these relations, and confirms the lack of correlation.} These findings suggest the possibility that centrist parties in the economic dimension may be single-issue' in any of the aforementioned three policy dimensions, or that centrist parties in the social dimension may be single-issue' in the dimension of European integration.\footnote{These results are consistent with the extensive literature that has examined the issue of the dimensionality of party spaces, particularly questioning the extent to which more than one dimension is relevant \citep{BenoitLaver2006, Marksetalt2006, Albright2010, Stoll2011, BenoitLaver2012, RovnyEdwards2012, BakkerBakkerPolk2012, RovnyPolk2019}. While there is no consensus in this literature on how many dimensions are necessary to analyze political issues without losing significant information,  there is agreement on the correlation between many dimensions, especially within the European context. Some dimensions, like European Integration and immigration, are considered exceptional \citep[in this regard, see][]{Marksetalt2006, RovnyEdwards2012, OtjesKatsanidou2017, ToshkovKrouwel2022}.
 }
 
To investigate whether the variation in the dimensions of European integration, urban/rural interests, and anti-Islam rhetoric plays a crucial role in explaining the correlation between centrism and blurriness in the economic and/or social dimensions (as reported in Section \ref{mainresults}), we introduce an interaction term in Equation (\ref{eqalt}). This interaction term captures the relationship between the level of centrism in the economic or social dimension and the level of centrism in each of these three additional dimensions. Our hypothesis is that if centrist parties adopt a single-issue approach in any of these additional dimensions, and if it is due to this single-issue focus that they exhibit blurriness in the economic or social dimensions, then the positive correlation between centrism and blurriness found in Section \ref{mainresults} should be stronger when these parties demonstrate less centrism on the issues they prioritize. In other words, the more extreme the single-issue stance of parties in these dimensions, the more blurred their positions in the economic or social dimensions would need to be to effectively attract voters.

Table \ref{tab2_singleissue} presents the results of the specified model, where columns (1) and (2) focus on European integration, columns (3) and (4) on urban/rural interests, and columns (5) and (6) on anti-Islam rhetoric. Importantly, among all the specifications, only one interaction term is statistically significant, and surprisingly, it exhibits a positive sign, contrary to what the single-issue hypothesis would suggest.\footnote{Specifically, the significant coefficient corresponds to the interaction between the level of centrism in the economic dimension and the level of centrism in the European integration dimension (column (1)), indicating that the more centrist (or neutral) the orientation of party leadership towards European integration, the more blurred centrist parties' positions are in the economic dimension.}

Additionally, we consider another relevant characteristic of political parties, for which we have data, which is their emphasis on reducing political corruption. It is plausible that certain single-issue political parties prioritize the fight against corruption and, in an effort to attract votes, strategically blur their economic and social positions. In columns (7) and (8) of Table \ref{tab2_singleissue}, we empirically analyze the plausibility of this hypothesis using data on how salient corruption is on the platform of each political party. The hypothesis is that the more single-issue the political parties are in the 'dimension' of corruption (i.e., the more significant reducing political corruption is for the parties), the more correlated their centrism in the economic and social dimensions should be with their blurriness. However, the interaction term in this last specification is not statistically significant for either the economic dimension or the social dimension.

The results in this subsection provide evidence against the single-issue hypothesis, as the ideological position of political parties across the dimensions examined does not seem to explain the correlation between centrism and blurriness in the economic and/or social dimensions reported in the previous section. Although this evidence is insufficient to completely rule out the single-issue hypothesis (since political parties might still be `single-issue' in dimensions for which data is unavailable, or this hypothesis could apply to a subset of our sample), considering the diverse range of dimensions included in this analysis, we find this explanation less plausible.\footnote{An important study related to ours, conducted by \cite{Rovny2012}, empirically investigates this issue. Using CHES data from 2006, the author finds correlations suggesting that parties further from the center on a particular dimension emphasize that dimension while also adopting centrist and blurred positions on other dimensions. Table \ref{tab_singleissuealt} in Appendix \ref{tables_A} extends this analysis to more periods and employs econometric specifications that include country-year and party fixed effects. Notably, the findings of \cite{Rovny2012} do not appear robust when these new specifications are employed, particularly when using our preferred outcome measure of party blurriness (columns (1) and (3)). While similar results to \cite{Rovny2012} are obtained for the dimension of social and cultural values when the outcome is the standard deviation of expert assessments regarding the ideology of each party (column (4)), the estimates are not jointly statistically different from zero. This suggests that while the hypothesis of \citeauthor{Rovny2012} plausibly explains the behavior of some 'single-issue' political parties, it may not adequately capture the possibly more general effect found in Section \ref{mainresults}.} 

\subsection{Post-electoral policy bias mitigation}

Another potential alternative explanation for the results in Section \ref{mainresults} is based on the idea that certain policy-relevant information is only revealed to politicians after elections. As a result, political parties may strategically blur their positions to adapt policies based on new information discovered while in office \citep{AragonesNeeman2000, Kartiketalt2017}. If centrist political parties are more concerned than extremist parties about policy adaptability, their platforms might exhibit higher levels of blurriness, potentially explaining the correlations observed in Section \ref{mainresults}.

Empirically assessing this hypothesis is challenging, primarily because it is difficult to identify ex ante which political parties prioritize policy adaptability. As an alternative approach, we explore a crucial implication of this hypothesis: if the correlation between centrism and blurriness reported in Section \ref{mainresults} is indeed driven by parties' concerns about policy adaptability, then this correlation should be stronger when such adaptability becomes more relevant. Economic crises represent such critical moments, where uncertainty regarding the effectiveness of policies is heightened, and it is plausible that parties particularly concerned about policy adaptability would be especially blurred during such crises.\footnote{In this particular hypothesis, our focus is on economic crises. However, an alternate perspective could posit that situations characterized by substantial GDP growth variance might also qualify as critical moments wherein parties with a distinct emphasis on policy adaptability could exhibit heightened ambiguity. Should this proposition hold true, one could infer that the empirical evidence presented in Section \ref{mechanisms1} would be congruent with this hypothesis, thereby diminishing the rationale for favouring the explanation presented in that section. Nevertheless, despite acknowledging the plausibility of this alternate hypothesis, there is a crucial aspect regarding the evidence presented in Section \ref{mechanisms1} which appears less congruent with the hypothesis currently under scrutiny:  that centrist parties are more ambiguous not only within countries undergoing pronounced GDP growth variation in the preceding year but notably when they did not hold power during that same year. If high GDP growth variance itself engenders increased ambiguity in centrist political parties, why would it be so important that these parties were not in power during the occurrence of such events? This aspect explains our inclination to prefer the hypothesis advanced in Section \ref{mechanisms1} over the alternative discussed in this subsection.}

Table \ref{tab3_postpolicybias} presents the results of a specification in which we include an interaction term between the level of centrism of each party in the economic or social dimension and a dummy variable that equals 1 if the real GDP growth of the country (contemporaneous and one-year lagged) falls below the median, and 0 if it exceeds the median. Notably, none of the interaction coefficients are statistically different from zero.\footnote{As for the estimates provided in Table \ref{tab4_uncertainty}, we derived these estimates using data on real GDP growth from the IMF.  In Appendix \ref{tables_A}, Panel B of Table \ref{tabA6_crisisrob} reports estimates similar to those in Table \ref{tab3_postpolicybias}, but using a continuous measure of real GDP growth and the proportion of years in the past in which each country experienced negative GDP growth. The results are similar to those in Table \ref{tab3_postpolicybias}: in neither specification is the interaction coefficient statistically different from zero.}

To address concerns about the GDP growth measure's ability to capture economic crises accurately, we conduct a robustness analysis in Panel B of Table \ref{tabA6_crisisrob} in Appendix \ref{tables_A}. We use data on financial crises from \cite{LaevenValencia2020} to examine whether the relationship between centrism and blurriness is more pronounced in countries that have experienced a greater number of financial crises in the past (e.g., banking crises or currency crises). However, even in this analysis, the interaction coefficients remain statistically insignificant, consistent with the results in Table \ref{tab3_postpolicybias}.

In summary, the results of Tables \ref{tab3_postpolicybias} and \ref{tabA6_crisisrob} do not provide support for the post-electoral policy bias mitigation hypothesis. While this does not entirely rule out the possibility of the hypothesis holding true in specific contexts, there is currently insufficient evidence in its favor.


\section{Conclusion}\label{Conclusion}

In this study, we have explored the relationship between ideological ambiguity and the political positioning of European political parties. Our key finding is a robust non-monotonic (inverted U-shaped) relationship between party ideology and ideological blurriness, observed consistently across all dimensions we analyzed. The results indicate that party ideology influences the degree of blurriness.

To explain this pattern, we have proposed a simple and novel theory based on voters perceiving centrist political parties as less risky than extremist parties. This perception arises because extremist parties often propose untested and unconventional policies, while centrist parties present more established and moderate policies. Given that the median voter is generally risk-averse, centrist parties have a natural advantage over extremists. Consequently, centrist parties are incentivized to strategically employ ambiguity to approach the median voter's ideal point more closely and secure electoral victories, given the lower relative cost of ambiguity for them. Our explanation finds support in additional evidence, including data on electoral outcomes and economic indicators from the countries where the studied parties operate. Furthermore, we have explored the empirical plausibility of two alternative explanations for the observed non-monotonic relationship between ideology and political position. However, our analysis does not support these alternative explanations; instead, the evidence seems to contradict them.

Looking ahead, there are several promising avenues for future research. One area is to further improve the identification of causal effects for the reported patterns. Additionally, investigating whether these patterns hold true for political parties in other regions, such as Latin America and Africa, could provide valuable insights. Moreover, delving deeper into the empirical examination of theories explaining these patterns, particularly identifying scenarios where certain theories apply better than others, presents an intriguing direction for future study. Lastly, exploring strategies or policies that limit or counter political parties' use of ambiguity, under the assumption that such ambiguity might be detrimental to democracy, could offer valuable insights for improving democratic governance.


\hbox {} \newpage
\bibliographystyle{aer}
\bibliography{msEG}


\hbox {} \newpage


\section*{Figures and Tables}

\begin{table}[H]
{ 
\renewcommand{\arraystretch}{0.5} 
\setlength{\tabcolsep}{1pt}
\captionsetup{font={normalsize,bf}}
\caption {Party position and blurriness: baseline results}  \label{tab1_baseline}
\begin{center}  
\small
\vspace{-0.5cm}\begin{tabular}{lcccccccc}
\hline\hline  \addlinespace[0.15cm]
 & \multicolumn{8}{c}{Dep. variable: Blurriness of each party's position} \\\cmidrule[0.2pt](l){2-9}\addlinespace[0.10cm] 
& (1)& (2) & (3)& (4)& (5)&(6)&(7)&(8) \\  \addlinespace[0.1cm] \hline \addlinespace[0.15cm] 
        
         \multicolumn{1}{l}{\emph{\underline{Panel A}:}}      & \multicolumn{7}{c}{Economic issues} \\\cmidrule[0.2pt](l){2-9}
\input{table1A.tex}
\addlinespace[0.1cm] \hline \addlinespace[0.15cm] 
         \multicolumn{1}{l}{\emph{\underline{Panel B}:}}      & \multicolumn{7}{c}{Social and cultural values} \\\cmidrule[0.2pt](l){2-9}
\input{table1B.tex}
\addlinespace[0.15cm]\hline\hline\addlinespace[0.15cm]    
  Country fixed effects & N & Y & - & - & - & - & -& -\\      
  Year fixed effects & N & Y & - & - & -& -& - & -\\         
  Country $\times$ Year fixed effects & N & N & Y & Y & Y& Y& Y& Y   \\       
 Political party fixed effects & N & N & N & Y & N & Y & N& Y\\    
      \addlinespace[0.15cm] \hline\hline         
\multicolumn{9}{p{16.7cm}}{\footnotesizes{\textbf{Notes:} 
Columns (1) to (4) report fixed effects OLS estimates  from Eq (\ref{eqbaseline}). Columns (5) and (6) report fixed effects OLS estimates from Eq (\ref{eqalt}). Columns (7) and (8) report fixed effects OLS estimates  from a monotonic relationship between a party's position and blurriness. The dependent variable in all columns is each party's blurriness on economic issues (Panel A) and social and cultural values (Panel B).  The sample is limited to the years 2017 and 2019.  In this sample, the average party position on economic issues is  4.916 (with s.d. 2.167), the average position on social and cultural values is 5.004 (with s.d. 2.724),  the average level of blurriness in economic issues is 3.675 (with s.d.  1.731) and  the average level of blurriness on on social and cultural values is 2.824 (with s.d. 1.548).  Robust standard errors  clustered by political party are in parentheses. * denotes results are statistically significant at the 10\% level, ** at the 5\% level, and *** at the 1\% level.} } \\
\end{tabular}
\end{center}
}
\end{table}


\begin{table}[H]
{ 
\renewcommand{\arraystretch}{0.9} 
\setlength{\tabcolsep}{2pt}
\captionsetup{font={normalsize,bf}}
\caption {Party position and blurriness: uncertainty of extremes}  \label{tab4_uncertainty}
\begin{center}  
\small
\vspace{-0.5cm}\begin{tabular}{lcccccccc}
\hline\hline  \addlinespace[0.15cm]
      & \multicolumn{6}{c}{Dep. variable: Blurriness on} \\\cmidrule[0.2pt](l){2-7}\addlinespace[0.10cm]    
     & \multicolumn{3}{c}{Economic issues}& \multicolumn{3}{c}{Social values}  \\\cmidrule[0.2pt](l){2-4}\cmidrule[0.2pt](l){5-7}
          & (1)& (2) & (3)& (4) & (5)& (6)  \\  \addlinespace[0.1cm] \cmidrule[0.2pt](l){2-4}\cmidrule[0.2pt](l){5-7} \addlinespace[0.10cm] 
\input{table4.tex}
\addlinespace[0.15cm]\hline\hline\addlinespace[0.15cm]    
  Country $\times$ Year fixed effects & Y & Y & Y   & Y & Y & Y    \\  
   Political party fixed effects & N & N & Y & N  & N & Y    \\        
      \addlinespace[0.15cm] \hline\hline         
\multicolumn{7}{p{16cm}}{\footnotesizes{\textbf{Notes:} 
All columns report fixed effects OLS estimates from an specification in which we introduce into Eq. (\ref{eqalt}) interaction terms between the level of economic or social centrism of each party and (i) the one-year lagged GDP growth variance of the country in which each respective party operates, and (ii) a dummy variable denoting whether the party held no governmental position during the same year. The dependent variable in columns (1) to (3) is each party's blurriness on economic issues. The dependent variable in columns (4) to (6) is each party's blurriness on social and cultural values.  The sample is limited to the years 2017 and 2019.  Robust standard errors  clustered by political party are in parentheses. * denotes results are statistically significant at the 10\% level, ** at the 5\% level, and *** at the 1\% level.} } \\
\end{tabular}
\end{center}
}
\end{table}


\begin{table}[H]
{ 
\renewcommand{\arraystretch}{0.5} 
\setlength{\tabcolsep}{2pt}
\captionsetup{font={normalsize,bf}}
\caption {Party position and blurriness:  role of European integration, urban/rural interests, anti-Islam rhetoric and corruption salience}  \label{tab2_singleissue}
\begin{center}  
\small
\vspace{-0.5cm}\begin{tabular}{@{}rcccccccc@{}}
\hline\hline  \addlinespace[0.15cm]
  \multicolumn{1}{l}{\hspace{0.5em}}     & \multicolumn{8}{c}{Dep. variable: Blurriness of each party's position} \\\cmidrule[0.2pt](l){2-9}\addlinespace[0.10cm] 
& (1)& (2) & (3)& (4)& (5)&(6)&(7)&(8) \\  \addlinespace[0.1cm] \hline \addlinespace[0.15cm] 
                                \multicolumn{1}{l}{Centrism on:\hspace{0.3em}}      &  \\\addlinespace[0.3cm]     
\input{table2.tex}
\addlinespace[0.15cm]\hline\hline\addlinespace[0.15cm]     
  Country $\times$ Year fixed effects & Y & Y  & Y & Y & Y& Y& Y& Y   \\       
      \addlinespace[0.15cm] \hline\hline         
\multicolumn{9}{p{16.8cm}}{\footnotesizes{\textbf{Notes:} 
All columns report fixed effects OLS estimates from an specification in which we introduce into Eq. (\ref{eqalt}) an interaction term between the level of economic or social centrism of each party and the position of each party regarding the European integration, urban/rural interests, anti-Islam rhetoric and the salience of corruption. The dependent variable in all columns is each party's blurriness on each party's issue. The sample is limited to the years 2017 and 2019.  Robust standard errors  clustered by political party are in parentheses. * denotes results are statistically significant at the 10\% level, ** at the 5\% level, and *** at the 1\% level.} } \\
\end{tabular}
\end{center}
}
\end{table}


\begin{table}[H]
{ 
\renewcommand{\arraystretch}{0.9} 
\setlength{\tabcolsep}{0.6pt}
\captionsetup{font={normalsize,bf}}
\caption {Party position and blurriness: role of GDP growth}  \label{tab3_postpolicybias}
\begin{center}  
\small
\vspace{-0.5cm}\begin{tabular}{lcccccccc}
\hline\hline  \addlinespace[0.15cm]
& (1)& (2) & (3)& (4) & (5)& (6) & (7)& (8) \\  \addlinespace[0.1cm] \hline \addlinespace[0.15cm] 
      & \multicolumn{8}{c}{Dep. variable: Blurriness on} \\\cmidrule[0.2pt](l){2-9}\addlinespace[0.10cm]    
     & \multicolumn{4}{c}{Economic issues}& \multicolumn{4}{c}{Social values}  \\\cmidrule[0.2pt](l){2-5}\cmidrule[0.2pt](l){6-9}
     
\input{table3.tex}
\addlinespace[0.15cm]\hline\hline\addlinespace[0.15cm]    
  Country $\times$ Year fixed effects & Y & Y & Y & Y  & Y & Y & Y & Y   \\  
   Political party fixed effects & N & Y & N& Y  & N & Y & N& Y  \\         
      \addlinespace[0.15cm] \hline\hline         
\multicolumn{9}{p{16.5cm}}{\footnotesizes{\textbf{Notes:} 
All columns report fixed effects OLS estimates from an specification in which we introduce into Eq. (\ref{eqalt}) an interaction term between the level of economic or social centrism of each party and the GDP growth of each country.  The dependent variable in columns (1) to (4) is each party's blurriness on economic issues. The dependent variable in columns (5) to (8) is each party's blurriness on social and cultural values. The sample is limited to the years 2017 and 2019.  Robust standard errors  clustered by political party are in parentheses. * denotes results are statistically significant at the 10\% level, ** at the 5\% level, and *** at the 1\% level.} } \\
\end{tabular}
\end{center}
}
\end{table}


\hbox {} \newpage

\appendix

\section{Appendix}
\subsection{Additional Tables}\label{tables_A}

\setcounter{table}{0}
\setcounter{figure}{0}
\renewcommand{\thefigure}{\Alph{section}\arabic{figure}}
\renewcommand{\thetable}{\Alph{section}\arabic{table}}

\begin{table}[H]
{ 
\renewcommand{\arraystretch}{0.5} 
\setlength{\tabcolsep}{1pt}
\captionsetup{font={normalsize,bf}}
\caption {Party position and blurriness: baseline results at the expert level}  \label{tableA1}
\begin{center}  
\small
\vspace{-0.5cm}\begin{tabular}{lcccccccc}
\hline\hline  \addlinespace[0.15cm]
 & \multicolumn{8}{c}{Dep. variable: Blurriness of each party's position} \\\cmidrule[0.2pt](l){2-9}\addlinespace[0.10cm] 
& (1)& (2) & (3)& (4)& (5)&(6)&(7)&(8) \\  \addlinespace[0.1cm] \hline \addlinespace[0.15cm] 
        
         \multicolumn{1}{l}{\emph{\underline{Panel A}:}}      & \multicolumn{7}{c}{Economic issues} \\\cmidrule[0.2pt](l){2-9}
\input{tableA1A.tex}
\addlinespace[0.1cm] \hline \addlinespace[0.15cm] 
         \multicolumn{1}{l}{\emph{\underline{Panel B}:}}      & \multicolumn{7}{c}{Social and cultural values} \\\cmidrule[0.2pt](l){2-9}
\input{tableA1B.tex}
\addlinespace[0.15cm]\hline\hline\addlinespace[0.15cm]    
  Country fixed effects & Y & N & Y & - & - & - & -& -\\      
  Year fixed effects & Y & N & Y & - & -& -& - & -\\         
  Country $\times$ Year fixed effects & Y & N & N & Y & Y& Y& Y& Y   \\       
 Political party fixed effects & N & N & N & Y & N & Y & N& Y\\    
      \addlinespace[0.15cm] \hline\hline         
\multicolumn{9}{p{16.8cm}}{\footnotesizes{\textbf{Notes:} 
Columns (1) to (4) report fixed effects OLS estimates for estimates from Eq (\ref{eqbaseline}). Columns (5) and (6) report fixed effects OLS estimates for estimates from Eq (\ref{eqalt}). Columns (7) and (8) report fixed effects OLS estimates for estimates from a monotonic relationship between a party's position and blurriness. The dependent variable in all columns is each party's blurriness on economic issues (Panel A) and social and cultural values (Panel B).  The sample is limited to the years 2017 and 2019.  Robust standard errors  clustered by political party are in parentheses. * denotes results are statistically significant at the 10\% level, ** at the 5\% level, and *** at the 1\% level.} } \\
\end{tabular}
\end{center}
}
\end{table}



\begin{table}[H]
{ 
\renewcommand{\arraystretch}{0.5} 
\setlength{\tabcolsep}{10pt}
\captionsetup{font={normalsize,bf}}
\caption {Party position and blurriness: alternative definition of centrism}  \label{tab_centrismalt}
\begin{center}  
\small
\vspace{-0.5cm}\begin{tabular}{lcccccccc}
\hline\hline  \addlinespace[0.15cm]
 & \multicolumn{4}{c}{Dep. variable: Blurriness of each party's position} \\\cmidrule[0.2pt](l){2-5}\addlinespace[0.10cm] 
& (1)& (2) & (3)& (4) \\  \addlinespace[0.1cm] \hline \addlinespace[0.15cm] 
        
   & \multicolumn{2}{c}{Economic issues}  & \multicolumn{2}{c}{Social and cultural values} \\\cmidrule[0.2pt](l){2-3}\cmidrule[0.2pt](l){4-5}
\input{tableAcentrismalt.tex}

\addlinespace[0.15cm]\hline\hline\addlinespace[0.15cm]    
  Country $\times$ Year fixed effects & Y & Y & Y &Y  \\       
 Political party fixed effects & N & Y & N & Y \\    
      \addlinespace[0.15cm] \hline\hline         
\multicolumn{5}{p{14cm}}{\footnotesizes{\textbf{Notes:} 
All columns report fixed effects OLS estimates for estimates from Eq (\ref{eqalt}), using the alternative definition of centrism described in footnote 13. The dependent variable in all columns is each party's blurriness on economic issues (columns (1) and (2)) and social and cultural values (columns (3) and (4).   The sample is limited to the years 2017 and 2019.  Robust standard errors  clustered by political party are in parentheses. * denotes results are statistically significant at the 10\% level, ** at the 5\% level, and *** at the 1\% level.} } \\
\end{tabular}
\end{center}
}
\end{table}


\begin{figure}[H]
     \centering
             \caption{Party position on economic and social issues and standard deviation (SD) in expert opinions}
        \label{fig_blurrpositionSD}
     \begin{subfigure}[b]{0.51\textwidth}
         \centering
         \includegraphics[width=\textwidth]{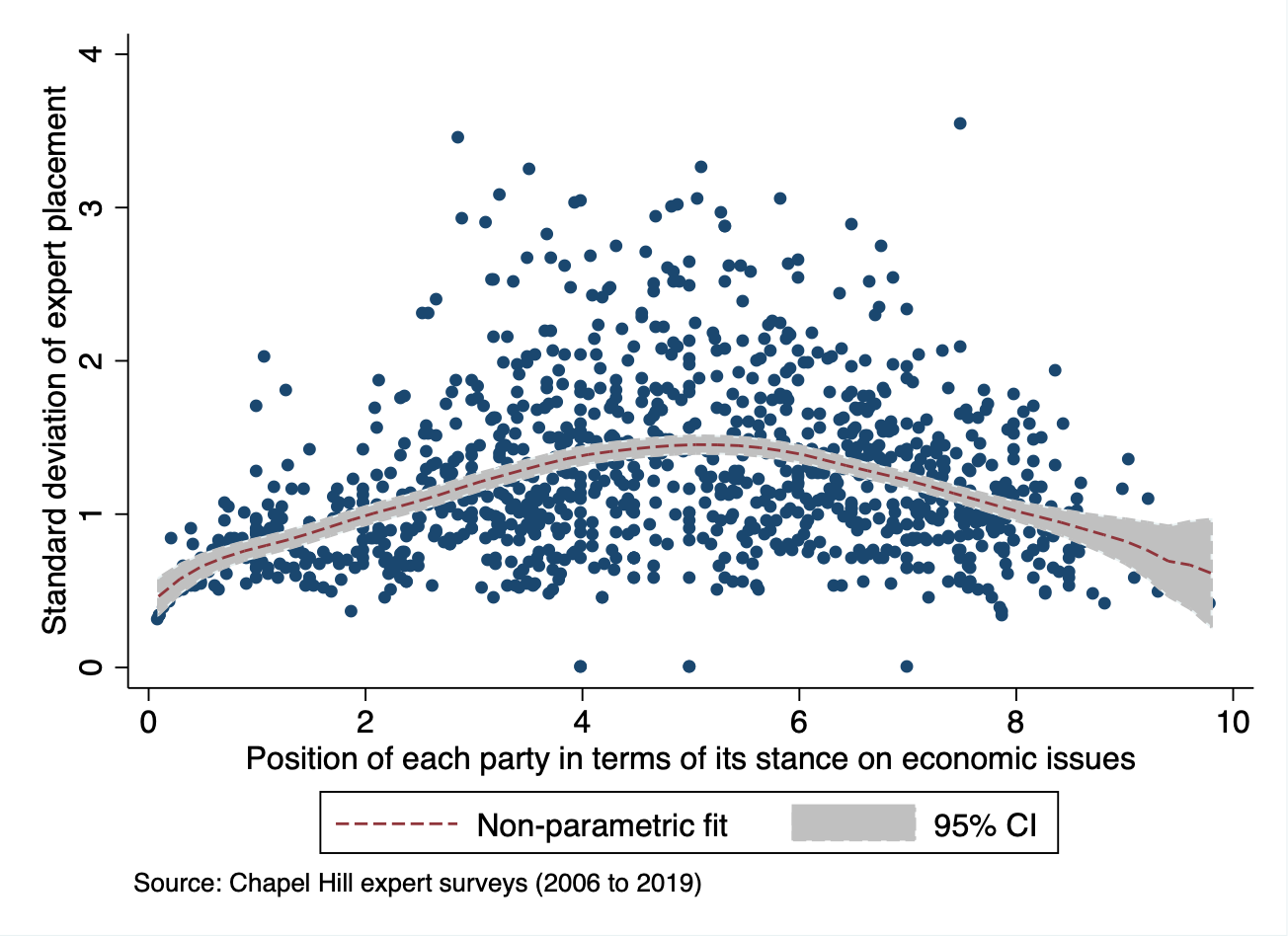}
         \caption{Economic issues}
         \label{fig_blurrpositionSD_a}
     \end{subfigure}
     \hspace{-1cm}.
     \begin{subfigure}[b]{0.51\textwidth}
         \centering
         \includegraphics[width=\textwidth]{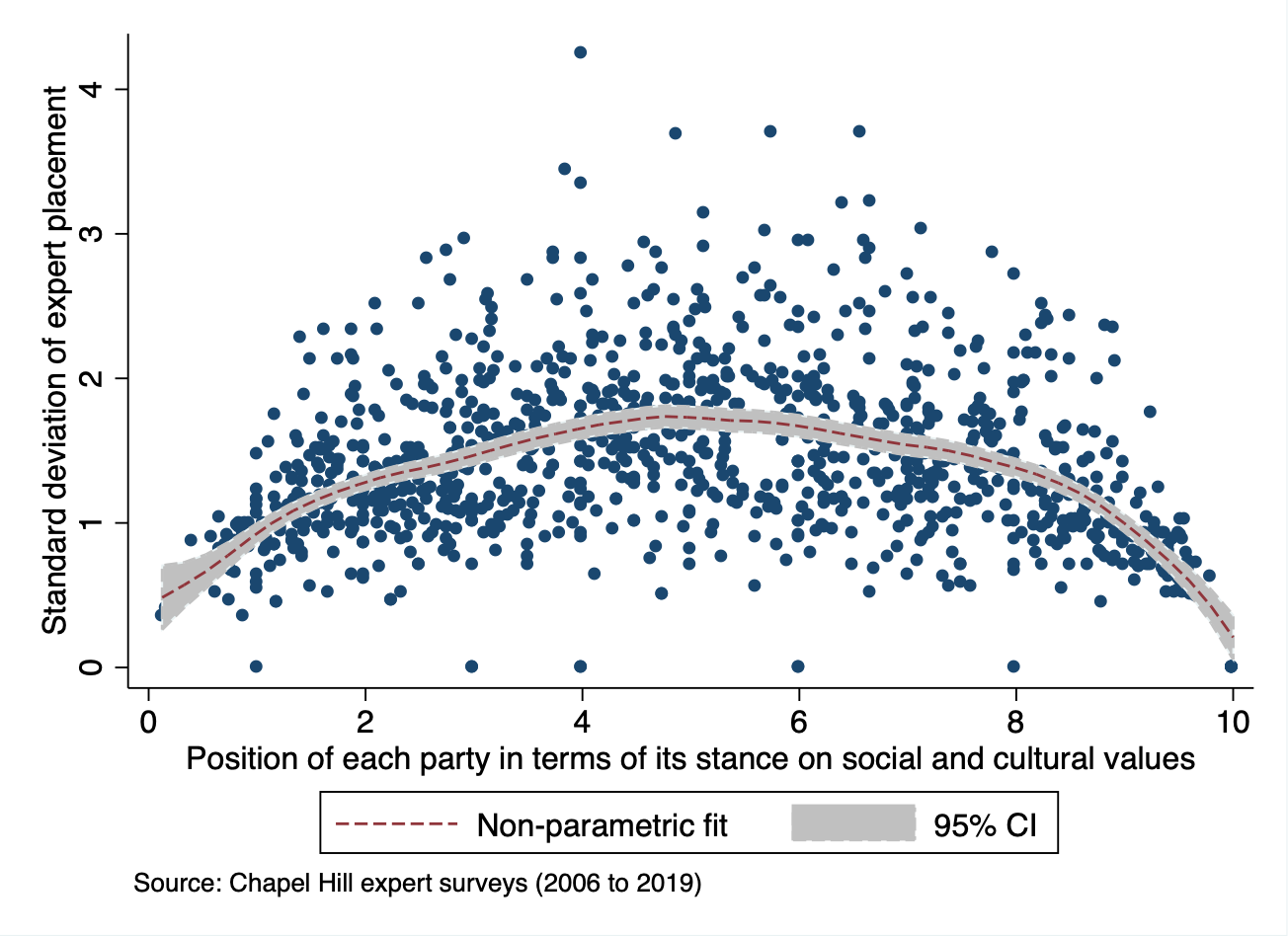}
         \caption{Social values}
         \label{fig_blurrpositionSD_b}
     \end{subfigure}
\end{figure}


\begin{table}[H]
{ 
\renewcommand{\arraystretch}{0.5} 
\setlength{\tabcolsep}{1pt}
\captionsetup{font={normalsize,bf}}
\caption {Party position and standard deviation (SD) in expert opinions: baseline results}  \label{tab_baselineSD}
\begin{center}  
\small
\vspace{-0.5cm}\begin{tabular}{lcccccccc}
\hline\hline  \addlinespace[0.15cm]
 & \multicolumn{8}{c}{Dep. variable: Blurriness of each party's position} \\\cmidrule[0.2pt](l){2-9}\addlinespace[0.10cm] 
& (1)& (2) & (3)& (4)& (5)&(6)&(7)&(8) \\  \addlinespace[0.1cm] \hline \addlinespace[0.15cm] 
        
   & \multicolumn{4}{c}{Economic issues}  & \multicolumn{4}{c}{Social and cultural values} \\\cmidrule[0.2pt](l){2-5}\cmidrule[0.2pt](l){6-9}
\input{tableA2f.tex}

\addlinespace[0.15cm]\hline\hline\addlinespace[0.15cm]    
  Country fixed effects & N & Y & - & - & N & Y & -& -\\      
  Year fixed effects & N & Y & - & - & N& Y& - & -\\         
  Country $\times$ Year fixed effects & N & N & Y & Y & N& N& Y& Y   \\       
 Political party fixed effects & N & N & N & Y & N & N & N& Y\\    
      \addlinespace[0.15cm] \hline\hline         
\multicolumn{9}{p{17.5cm}}{\footnotesizes{\textbf{Notes:} 
All columns report fixed effects OLS estimates for estimates from Eq (\ref{eqbaseline}). The dependent variable  is each party's standard deviation (SD) in expert opinions on economic issues (columns (1) to (4)) and social and cultural values (columns (5) to (8).  The sample includes the years 2006, 2010, 2014, 2017 and 2019.  Robust standard errors  clustered by political party are in parentheses. * denotes results are statistically significant at the 10\% level, ** at the 5\% level, and *** at the 1\% level.} } \\
\end{tabular}
\end{center}
}
\end{table}


\begin{table}[H]
{ 
\renewcommand{\arraystretch}{0.5} 
\setlength{\tabcolsep}{1pt}
\captionsetup{font={normalsize,bf}}
\caption {Party position and blurriness: robustness to inclusion of additional controls}  \label{tableA2}
\begin{center}  
\small
\vspace{-0.5cm}\begin{tabular}{lcccccccc}
\hline\hline  \addlinespace[0.15cm]
 & \multicolumn{8}{c}{Dep. variable: Blurriness of each party's position} \\\cmidrule[0.2pt](l){2-9}\addlinespace[0.10cm] 
& (1)& (2) & (3)& (4)& (5)& (6)& (7) & (8) \\  \addlinespace[0.1cm] \hline \addlinespace[0.15cm] 
        
         \multicolumn{1}{l}{\emph{\underline{Panel A}:}}      & \multicolumn{8}{c}{Economic issues} \\\cmidrule[0.2pt](l){2-9}
\input{tableA2A.tex}
\addlinespace[0.1cm] \hline \addlinespace[0.15cm] 
         \multicolumn{1}{l}{\emph{\underline{Panel B}:}}      & \multicolumn{8}{c}{Social and cultural values} \\\cmidrule[0.2pt](l){2-9}
\input{tableA2B.tex}
\addlinespace[0.15cm]\hline\hline\addlinespace[0.15cm]     
  Control for internal dissent & Y & N & N & Y & Y  & N & N & Y   \\       
  Control for party age & N & Y & N & Y & N  & Y & N& Y   \\       
  Control for  government status  & N & N & Y & Y & N  & N & Y& Y   \\       
\addlinespace[0.15cm]\hline\hline\addlinespace[0.15cm]     
  Country $\times$ Year fixed effects & Y & Y & Y & Y & Y  & Y & Y& Y   \\       
      \addlinespace[0.15cm] \hline\hline         
\multicolumn{9}{p{16.7cm}}{\footnotesizes{\textbf{Notes:} 
Columns (1) to (4) report fixed effects OLS estimates for estimates from Eq (\ref{eqbaseline}). Columns (5) and (8) report fixed effects OLS estimates for estimates from Eq (\ref{eqalt}). The dependent variable in all columns is each party's blurriness on economic issues (Panel A) and social and cultural values (Panel B).  The sample is limited to the years 2017 and 2019.  Robust standard errors  clustered by political party are in parentheses. * denotes results are statistically significant at the 10\% level, ** at the 5\% level, and *** at the 1\% level.} } \\
\end{tabular}
\end{center}
}
\end{table}

\newpage


\begin{table}[H]
{ 
\renewcommand{\arraystretch}{0.8} 
\setlength{\tabcolsep}{1pt}
\captionsetup{font={normalsize,bf}}
\caption {Party position and blurriness:  IV estimates}  \label{tableA3}
\begin{center}  
\small
\vspace{-0.5cm}\begin{tabular}{lcccccccc}
\hline\hline  \addlinespace[0.15cm]
 & \multicolumn{8}{c}{Dep. variable: Blurriness of each party's position} \\\cmidrule[0.2pt](l){2-9}\addlinespace[0.10cm] 
& (1)& (2) & (3)& (4)& (5)&(6)& (7)&(8)  \\  \addlinespace[0.1cm] \hline \addlinespace[0.15cm] 
         \multicolumn{1}{l}{\emph{\underline{Panel A}:}}      & \multicolumn{8}{c}{Economic issues} \\\cmidrule[0.2pt](l){2-9}
\input{tableA3A.tex}
\addlinespace[0.1cm] \hline \addlinespace[0.15cm] 
         \multicolumn{1}{l}{\emph{\underline{Panel B}:}}      & \multicolumn{8}{c}{Social and cultural values} \\\cmidrule[0.2pt](l){2-9}
\input{tableA3B.tex}
\addlinespace[0.15cm]\hline\hline\addlinespace[0.15cm]    
  Country fixed effects  & N & Y & - &  - & -& -&  - & - \\      
  Year fixed effects  & N & Y & - &  -& - & -&  - & - \\         
  Country $\times$ Year fixed effects  & N & N & Y & Y& Y& Y  & Y& Y \\       
 Political party fixed effects  & N & N & N  & Y & N& Y & N & Y\\    
      \addlinespace[0.15cm] \hline\hline         
\multicolumn{9}{p{16.5cm}}{\footnotesizes{\textbf{Notes:} 
All columns report second-stage IV estimates  using lags of position and centrism as instruments.  Columns (1) to (4) report the second-stage IV estimates from Eq (\ref{eqbaseline}). Columns (5) and (6) report the second-stage  IV estimates from Eq (\ref{eqalt}). Columns (7) and (8) report the second-stage  IV estimates from a monotonic relationship between a party's position and blurriness. The dependent variable in all columns is each party's blurriness on economic issues (Panel A) and social and cultural values (Panel B).  The sample is limited to the years 2017 and 2019.  Robust standard errors  clustered by political party are in parentheses. * denotes results are statistically significant at the 10\% level, ** at the 5\% level, and *** at the 1\% level.} } \\
\end{tabular}
\end{center}
}
\end{table}


\newpage

\begin{table}[H]
{ 
\renewcommand{\arraystretch}{0.9} 
\setlength{\tabcolsep}{0.6pt}
\captionsetup{font={normalsize,bf}}
\caption {Party position and blurriness: simple lags}  \label{tableA3aa}
\begin{center}  
\small
\vspace{-0.5cm}\begin{tabular}{lcccccccc}
\hline\hline  \addlinespace[0.15cm]
& (1)& (2) & (3)& (4) & (5)& (6) & (7)& (8) \\  \addlinespace[0.1cm] \hline \addlinespace[0.15cm] 
      & \multicolumn{8}{c}{Dep. variable: Blurriness on} \\\cmidrule[0.2pt](l){2-9}\addlinespace[0.10cm]    
     & \multicolumn{4}{c}{Economic issues}& \multicolumn{4}{c}{Social values}  \\\cmidrule[0.2pt](l){2-5}\cmidrule[0.2pt](l){6-9}
     
\input{tableA3aa.tex}
\addlinespace[0.15cm]\hline\hline\addlinespace[0.15cm]    
  Country $\times$ Year fixed effects & Y & Y & Y & Y  & Y & Y & Y & Y   \\  
   Political party fixed effects & N & Y & N& Y  & N & Y & N& Y  \\         
      \addlinespace[0.15cm] \hline\hline         
\multicolumn{9}{p{16.5cm}}{\footnotesizes{\textbf{Notes:} 
All columns report fixed effects OLS estimates from Eq. (\ref{eqalt}).  The dependent variable in columns (1) to (4) is each party's blurriness on economic issues. The dependent variable in columns (5) to (8) is each party's blurriness on social and cultural values. The sample is limited to the years 2017 and 2019.  Robust standard errors  clustered by political party are in parentheses. * denotes results are statistically significant at the 10\% level, ** at the 5\% level, and *** at the 1\% level.} } \\
\end{tabular}
\end{center}
}
\end{table}

\newpage

\begin{table}[H]
{ 
\renewcommand{\arraystretch}{0.9} 
\setlength{\tabcolsep}{2pt}
\captionsetup{font={normalsize,bf}}
\caption {Party position and blurriness: uncertainty of extremes (robustness to an alternative measure of GDP growth variance)}  \label{tab_uncertaintyGDPalt}
\begin{center}  
\small
\vspace{-0.5cm}\begin{tabular}{lcccccccc}
\hline\hline  \addlinespace[0.15cm]
      & \multicolumn{6}{c}{Dep. variable: Blurriness on} \\\cmidrule[0.2pt](l){2-7}\addlinespace[0.10cm]    
     & \multicolumn{3}{c}{Economic issues}& \multicolumn{3}{c}{Social values}  \\\cmidrule[0.2pt](l){2-4}\cmidrule[0.2pt](l){5-7}
          & (1)& (2) & (3)& (4) & (5)& (6)  \\  \addlinespace[0.1cm] \cmidrule[0.2pt](l){2-4}\cmidrule[0.2pt](l){5-7} \addlinespace[0.10cm] 
\input{table2altGDP.tex}
\addlinespace[0.15cm]\hline\hline\addlinespace[0.15cm]    
  Country $\times$ Year fixed effects & Y & Y & Y   & Y & Y & Y    \\  
   Political party fixed effects & N & N & Y & N  & N & Y    \\        
      \addlinespace[0.15cm] \hline\hline         
\multicolumn{7}{p{16cm}}{\footnotesizes{\textbf{Notes:} 
All columns report fixed effects OLS estimates from an specification in which we introduce into Eq. (\ref{eqalt}) interaction terms between the level of economic or social centrism of each party and (i) a dummy variable equal to one if the one-year lagged GDP growth variance of each country  is greater than the median of the distribution of GDP growth variances of all countries in that same year, and (ii) a dummy variable denoting whether the party held no governmental position during the same year. The dependent variable in columns (1) to (3) is each party's blurriness on economic issues. The dependent variable in columns (4) to (6) is each party's blurriness on social and cultural values.  The sample is limited to the years 2017 and 2019.  Robust standard errors  clustered by political party are in parentheses. * denotes results are statistically significant at the 10\% level, ** at the 5\% level, and *** at the 1\% level.} } \\
\end{tabular}
\end{center}
}
\end{table}


\begin{table}[H]
{ 
\renewcommand{\arraystretch}{0.8} 
\setlength{\tabcolsep}{10pt}
\captionsetup{font={normalsize,bf}}
\caption {Alternative test to the `single issue' hypothesis } \label{tab_singleissuealt}
\begin{center}  
\small
\vspace{-0.5cm}\begin{tabular}{lcccccccc}
\hline\hline  \addlinespace[0.15cm]
 & \multicolumn{4}{c}{Dep. variable: Blurriness of each party's position} \\\cmidrule[0.2pt](l){2-5}\addlinespace[0.10cm] 
& (1)& (2) & (3)& (4) \\  \addlinespace[0.1cm] \hline \addlinespace[0.15cm] 
        
   & \multicolumn{2}{c}{Economic issues}  & \multicolumn{2}{c}{Social values} \\\cmidrule[0.2pt](l){2-3}\cmidrule[0.2pt](l){4-5}
\input{tableA_singleissuealt.tex}

\addlinespace[0.15cm]\hline\hline\addlinespace[0.15cm]         
  Country $\times$ Year fixed effects & Y & Y & Y & Y   \\       
 Political party fixed effects & Y & Y & Y & Y \\    
      \addlinespace[0.14cm] \hline\hline         
\multicolumn{5}{p{14.5cm}}{\footnotesizes{\textbf{Notes:} 
All columns report fixed effects OLS estimates  from Eq (\ref{eqbaseline}). The dependent variable is each party's blurriness on economic issues (columns (1) and (2)) and social and cultural values (columns (3) and (4)). Columns (1) and (3) use our baseline measure of a party's ambiguity (the assessment of each expert regarding such ambiguity). Columns (2) and (4) measure a party's ambiguity using the standard deviation of each expert's assessment relative to the ideology of each party. The sample in columns (1) and (3) includes the years 2017 and 2019, while the sample in columns (2) and (4) encompasses the years 2006, 2010, 2014, 2017, and 2019.  Robust standard errors  clustered by political party are in parentheses. * denotes results are statistically significant at the 10\% level, ** at the 5\% level, and *** at the 1\% level.} } \\
\end{tabular}
\end{center}
}
\end{table}


\newpage

\begin{landscape}
\renewcommand{\arraystretch}{0.9} 
\setlength{\tabcolsep}{0.005cm}
\input{tableA5.tex}
\end{landscape}

\newpage

\begin{table}[H]
{ 
\renewcommand{\arraystretch}{0.0} 
\setlength{\tabcolsep}{0.2pt}
\captionsetup{font={normalsize,bf}}
\caption {Party position and blurriness: role of GDP growth and debt criss}  \label{tabA6_crisisrob}
\begin{center}  
\small
\vspace{-0.8cm}\begin{tabular}{lcccccccc}
\hline\hline  \addlinespace[0.15cm]
& (1)& (2) & (3)& (4) & (5)& (6) & (7)& (8) \\  \addlinespace[0.1cm] \hline \addlinespace[0.15cm] 
      & \multicolumn{8}{c}{Dep. variable: Blurriness on} \\\cmidrule[0.2pt](l){2-9}\addlinespace[0.10cm]    
     & \multicolumn{4}{c}{Economic issues}& \multicolumn{4}{c}{Social values}  \\\cmidrule[0.2pt](l){2-5}\cmidrule[0.2pt](l){6-9}
\addlinespace[0.15cm]\hline\hline\addlinespace[0.15cm]       
  \multicolumn{1}{l}{\emph{\underline{Panel A}:}}     \\\addlinespace[0.3cm]     
\input{tableA6a.tex}
\addlinespace[0.1cm]\hline\addlinespace[0.15cm]     
  Country $\times$ Year fixed effects & Y & Y & Y & Y  & Y & Y & Y & Y   \\  
   Political party fixed effects & N & Y & N& Y  & N & Y & N& Y  \\            
\addlinespace[0.1cm]\hline\hline\addlinespace[0.15cm]   
  \multicolumn{1}{l}{\emph{\underline{Panel B}:}}   \\\addlinespace[0.3cm]     
\input{tableA6b.tex}
\addlinespace[0.1cm]\hline\addlinespace[0.15cm]     
  Country $\times$ Year fixed effects & Y & Y & Y & Y  & Y & Y & Y & Y   \\  
   Political party fixed effects & N & N & N& N  & N & N & N& N  \\         
      \addlinespace[0.1cm] \hline\hline         
\multicolumn{9}{p{17.7cm}}{\footnotesizes{\textbf{Notes:} 
All columns report fixed effects OLS estimates from an specification in which we introduce into Eq. (\ref{eqalt}) an interaction term between the level of economic or social centrism of each party and the number of years with positive (or negative) GDP growth in each country (Panel A) and  the number of years with banking or debts crisis in each country (Panel B).  The dependent variable in columns (1) to (4) is each party's blurriness on economic issues. The dependent variable in columns (5) to (8) is each party's blurriness on social and cultural values. The sample is limited to the years 2017 and 2019.  Robust standard errors  clustered by political party are in parentheses. * denotes results are statistically significant at the 10\% level, ** at the 5\% level, and *** at the 1\% level.} } \\
\end{tabular}
\end{center}
}
\end{table}

\newpage


\subsection{Model for the main mechanism: proofs}\label{model_A}
  
In this section, we prove the main results outlined in Section \ref{secc_model}. Specifically, we delve into the strategic decision-making process of political parties concerning the adoption of an ambiguous stance. Our analysis shows that the equilibrium actions critically hinge on whether $k^{2}<\frac{3}{2}$ or $k^{2}>\frac{3}{2}$.

Using the payoffs that the median voter obtains in each possible scenario (as detailed in Section \ref{secc_model}), we proceed to examine the actions and best responses of each party in these scenarios. We begin by considering the scenario where the centrist party is ambiguous, and investigate what happens if the extremist party does not follow suit. If the extremist party also chooses to be ambiguous, as mentioned earlier, it will consistently end up losing. However, if the extremist party decides to specify a position, two distinct scenarios arise: (i) When $|-1|<|-\frac{2(k^{2}+1)}{5}|$, that is, when  $k^{2}>\frac{3}{2}$, the extremist party will defeat the centrist party with a probability of $\frac{3}{5}$. (ii) When $k^{2}<\frac{3}{2}$, the extremist party will only defeat the centrist party if the extremist party announces a position that aligns with the median voter's ideal point, and this occurs with a probability of $\frac{1}{5}$. \emph{Thus, regardless of $k$, the best response for the extremist party when the centrist party is ambiguous is to avoid ambiguity and instead specify a position.}

Let's shift our focus to the scenario where the extremist party adopts an ambiguous stance. If the centrist party also remains ambiguous, as mentioned earlier, the centrist party will always win. However, if the centrist party specifies a position, we encounter a situation similar to the one studied before: if $|-1|<|-\frac{2(l^{2}+k^{2}+1)}{7}|$, that is, if $k^{2}+l^{2}>\frac{5}{2}$, the centrist party will defeat the extremist party with a probability of $\frac{3}{5}$; and if $k^{2}+l^{2}<\frac{5}{2}$, the centrist party will defeat the extremist party only if the centrist party announced a position coincident with the median voter's ideal point, and this occurs with a probability of $\frac{1}{5}$. \emph{Thus, if the extremist party opts for an ambiguous stance, the best response for the centrist party is to also remain ambiguous, regardless of the specific values of $k$ or $l$.} 

Now, let's consider the scenario in which the extremist party specifies a position. If the centrist party also specifies a position, then, as previously mentioned, each party will have an equal probability of winning, with a chance of  $\frac{1}{2}$ for each. However, if the centrist party chooses instead to remain ambiguous, we encounter a situation similar to the one analyzed earlier: (i) When $|-1|<|-\frac{2(k^{2}+1)}{5}|$, meaning that $k^{2}>\frac{3}{2}$, the centrist party will defeat the extremist party with a probability of $\frac{2}{5}$. Notably, this probability is smaller than the $\frac{1}{2}$ probability when both parties specify positions. (ii) When $k^{2}<\frac{3}{2}$, the extremist party will only defeat the centrist party if the extremist party announces a position coinciding with the median voter's ideal point, and this occurs with a probability of $\frac{1}{5}$. \emph{Thus, when the extremist party specifies a position: (i) If $k^{2}>\frac{3}{2}$, the best response by the centrist party is to specify a position, as this yields better odds of winning than remaining ambiguous. (ii) If $k^{2}<\frac{3}{2}$, the best response by the centrist party is to remain ambiguous, given that this increases their chances of winning compared to specifying a position.}

Finally, let's now consider the scenario in which the centrist party specifies a position. If the extremist party also specifies a position, then, as previously mentioned, each party will have an equal probability of winning, with a chance of $\frac{1}{2}$ for each. However, if the extremist party chooses instead to remain ambiguous, we encounter a situation similar to the one studied before: (i) When $|-1|<|-\frac{2(l^{2}+k^{2}+1)}{7}|$, that is, if $k^{2}+l^{2}>\frac{5}{2}$, the centrist party will defeat the extremist party with a probability of $\frac{3}{5}$. (ii) When  $k^{2}+l^{2}<\frac{5}{2}$, the centrist party will defeat the extremist party only if the centrist party announced a position coincident with the median voter's ideal point, and this occurs with a probability of $\frac{1}{5}$. \emph{Thus, if the centrist party specifies a position: (i) If $k^{2}+l^{2}>\frac{5}{2}$, the best strategic response by the extremist party is to specify a position, as this yields better odds of winning than remaining ambiguous. (ii) If $k^{2}+l^{2}<\frac{5}{2}$, the best response by the extremist party is to remain ambiguous, given that this increases their chances of winning compared to specifying a position.}
        
        \medskip
        
    \noindent The previous results can be summarized as follows:
               \begin{itemize}
 \item[(i)] When $k^{2}<\frac{3}{2}$, the equilibrium outcome is that the centrist party remains ambiguous, while the extremist party adopts a non-ambiguous stance.
  \item[(ii)] When $k^{2}>\frac{3}{2}$, since this implies that $k^{2}+l^{2}>\frac{5}{2}$, the equilibrium outcome is that both the centrist and extremist parties choose not to be ambiguous.
    \item[(iii)] There is no equilibrium in which both parties are ambiguous or where only the extremist party is ambiguous.
\end{itemize}

\end{document}

%% file: table1A.tex
Position on economic issues&       1.848\sym{***}&       1.826\sym{***}&       1.844\sym{***}&       1.347\sym{**} &                     &                     &   0.072\sym{\dagger}&      -0.234         \\
                    &     (0.127)         &     (0.147)         &     (0.151)         &     (0.611)         &                     &                     &     (0.049)         &     (0.200)         \\
Position on economic issues sq.&      -0.186\sym{***}&      -0.184\sym{***}&      -0.186\sym{***}&      -0.142\sym{***}&                     &                     &                     &                     \\
                    &     (0.014)         &     (0.016)         &     (0.017)         &     (0.050)         &                     &                     &                     &                     \\
Centrism on economic issues&                     &                     &                     &                     &       0.810\sym{***}&       0.877\sym{***}&                     &                     \\
                    &                     &                     &                     &                     &     (0.084)         &     (0.293)         &                     &                     \\
Joint significance  &       0.000         &       0.000         &       0.000         &       0.005         &                     &                     &                     &                     \\
Vertex              &       4.981         &       4.965         &       4.957         &       4.735         &                     &                     &                     &                     \\
R-squared           &       0.303         &       0.494         &       0.517         &       0.895         &       0.443         &       0.897         &       0.264         &       0.888         \\
Observations        &         406         &         406         &         406         &         212         &         406         &         212         &         406         &         212         \\

%% file: table1B.tex
Position on social values&       1.722\sym{***}&       1.632\sym{***}&       1.664\sym{***}&       0.520         &                     &                     &       0.022         &      -0.200         \\
                    &     (0.095)         &     (0.092)         &     (0.091)         &     (0.415)         &                     &                     &     (0.032)         &     (0.207)         \\
Position on social values sq.&      -0.163\sym{***}&      -0.156\sym{***}&      -0.159\sym{***}&      -0.063\sym{**} &                     &                     &                     &                     \\
                    &     (0.009)         &     (0.009)         &     (0.009)         &     (0.030)         &                     &                     &                     &                     \\
Centrism on social values&                     &                     &                     &                     &       0.980\sym{***}&       0.702\sym{***}&                     &                     \\
                    &                     &                     &                     &                     &     (0.053)         &     (0.154)         &                     &                     \\
Joint significance  &       0.000         &       0.000         &       0.000         &       0.038         &                     &                     &                     &                     \\
Vertex              &       5.277         &       5.227         &       5.234         &       4.111         &                     &                     &                     &                     \\
R-squared           &       0.462         &       0.601         &       0.627         &       0.894         &       0.655         &       0.902         &       0.229         &       0.890         \\
Observations        &         406         &         406         &         406         &         214         &         406         &         214         &         406         &         214         \\

%% file: table4.tex
Centrism on economic issues&       0.836\sym{***}&       1.035\sym{***}&       2.799\sym{***}&                     &                     &                     \\
                    &     (0.089)         &     (0.351)         &     (0.718)         &                     &                     &                     \\
Centrism $\times$ GDP growth variance&   0.004\sym{\dagger}&      -0.084\sym{*}  &      -0.204\sym{***}&                     &                     &                     \\
                    &     (0.002)         &     (0.050)         &     (0.071)         &                     &                     &                     \\
\multirow{1}{*}{\shortstack{Centrism $\times$ GDP growth var $\times$ not in govt }}&                     &       0.089\sym{*}  &       0.228\sym{***}&                     &                     &                     \\
                    &                     &     (0.050)         &     (0.078)         &                     &                     &                     \\
Centrism $\times$ not in government&                     &      -0.191         &      -1.963\sym{***}&                     &                     &                     \\
                    &                     &     (0.370)         &     (0.738)         &                     &                     &                     \\
Centrism on social values&                     &                     &                     &       0.998\sym{***}&       1.184\sym{***}&       0.329         \\
                    &                     &                     &                     &     (0.060)         &     (0.179)         &     (0.784)         \\
Centrism $\times$ GDP growth variance&                     &                     &                     &       0.004         &       0.020         &      -0.050         \\
                    &                     &                     &                     &     (0.004)         &     (0.018)         &     (0.050)         \\
Centrism $\times$ GDP growth var $\times$ not in govt &                     &                     &                     &                     &      -0.014         &       0.053         \\
                    &                     &                     &                     &                     &     (0.019)         &     (0.052)         \\
Centrism $\times$ not in govt&                     &                     &                     &                     &      -0.240         &       0.386         \\
                    &                     &                     &                     &                     &     (0.197)         &     (0.807)         \\
GDP growth variance $\times$ not in government&                     &      -0.309\sym{*}  &      -0.745\sym{***}&                     &       0.067         &      -0.100         \\
                    &                     &     (0.180)         &     (0.238)         &                     &     (0.074)         &     (0.165)         \\
Party not in government&                     &       0.705         &       6.721\sym{**} &                     &       0.115         &      -0.762         \\
                    &                     &     (1.160)         &     (2.631)         &                     &     (0.569)         &     (2.515)         \\
R-squared           &       0.439         &       0.444         &       0.910         &       0.655         &       0.668         &       0.904         \\
Observations        &         349         &         349         &         194         &         348         &         348         &         196         \\

%% file: table2.tex
Economic issues     &      -0.148         &                     &       0.874\sym{***}&                     &       0.751\sym{***}&                     &       0.619\sym{***}&                     \\
                    &     (0.364)         &                     &     (0.271)         &                     &     (0.173)         &                     &     (0.161)         &                     \\
Social values       &                     &       0.755\sym{***}&                     &       0.775\sym{***}&                     &       0.815\sym{***}&                     &       0.936\sym{***}\\
                    &                     &     (0.158)         &                     &     (0.227)         &                     &     (0.138)         &                     &     (0.114)         \\
Economic issues $\times$ EU intg.&       0.464\sym{***}&                     &                     &                     &                     &                     &                     &                     \\
                    &     (0.152)         &                     &                     &                     &                     &                     &                     &                     \\
Social values $\times$  EU intg.&                     &       0.109         &                     &                     &                     &                     &                     &                     \\
                    &                     &     (0.078)         &                     &                     &                     &                     &                     &                     \\
European integration&      -0.831\sym{*}  &      -0.036         &                     &                     &                     &                     &                     &                     \\
                    &     (0.487)         &     (0.174)         &                     &                     &                     &                     &                     &                     \\
Economic issues $\times$ urban/rural&                     &                     &      -0.044         &                     &                     &                     &                     &                     \\
                    &                     &                     &     (0.089)         &                     &                     &                     &                     &                     \\
Social values $\times$ urban/rural&                     &                     &                     &       0.061         &                     &                     &                     &                     \\
                    &                     &                     &                     &     (0.070)         &                     &                     &                     &                     \\
Urban/rural interests&                     &                     &       0.305         &      -0.121         &                     &                     &                     &                     \\
                    &                     &                     &     (0.292)         &     (0.167)         &                     &                     &                     &                     \\
Economic issues $\times$ anti-Islam&                     &                     &                     &                     &       0.023         &                     &                     &                     \\
                    &                     &                     &                     &                     &     (0.103)         &                     &                     &                     \\
Social values $\times$ anti-Islam&                     &                     &                     &                     &                     &       0.084         &                     &                     \\
                    &                     &                     &                     &                     &                     &     (0.072)         &                     &                     \\
Anti-Islam rhetoric &                     &                     &                     &                     &      -0.108         &       0.005         &                     &                     \\
                    &                     &                     &                     &                     &     (0.320)         &     (0.192)         &                     &                     \\
Economic issues $\times$ corruption&                     &                     &                     &                     &                     &                     &   0.052\sym{\dagger}&                     \\
                    &                     &                     &                     &                     &                     &                     &     (0.033)         &                     \\
Social values $\times$ corruption&                     &                     &                     &                     &                     &                     &                     &       0.013         \\
                    &                     &                     &                     &                     &                     &                     &                     &     (0.027)         \\
Corruption salience &                     &                     &                     &                     &                     &                     &       0.003         &       0.019         \\
                    &                     &                     &                     &                     &                     &                     &     (0.108)         &     (0.052)         \\
R-squared           &       0.506         &       0.665         &       0.500         &       0.667         &       0.502         &       0.675         &       0.469         &       0.658         \\
Observations        &         406         &         406         &         274         &         275         &         269         &         270         &         406         &         406         \\

%% file: table3.tex
Centrism on economic issues&       0.926\sym{***}&       0.966\sym{***}&       0.873\sym{***}&       0.833\sym{**} &                     &                     &                     &                     \\
                    &     (0.108)         &     (0.262)         &     (0.102)         &     (0.279)         &                     &                     &                     &                     \\
Centrism $\times$ low GDP growth (t)&  -0.208\sym{\dagger}&      -0.215\sym{*}  &                     &                     &                     &                     &                     &                     \\
                    &     (0.139)         &     (0.106)         &                     &                     &                     &                     &                     &                     \\
Centrism $\times$ low GDP growth (t-1)&                     &                     &      -0.107         &       0.091         &                     &                     &                     &                     \\
                    &                     &                     &     (0.152)         &     (0.238)         &                     &                     &                     &                     \\
Centrism on social values&                     &                     &                     &                     &       0.996\sym{***}&       0.681\sym{***}&       0.993\sym{***}&       0.650\sym{***}\\
                    &                     &                     &                     &                     &     (0.066)         &     (0.134)         &     (0.081)         &     (0.172)         \\
Centrism $\times$ low GDP growth (t)&                     &                     &                     &                     &      -0.032         &       0.039         &                     &                     \\
                    &                     &                     &                     &                     &     (0.102)         &     (0.085)         &                     &                     \\
Centrism $\times$ low GDP growth (t-1)&                     &                     &                     &                     &                     &                     &      -0.025         &       0.077         \\
                    &                     &                     &                     &                     &                     &                     &     (0.100)         &     (0.109)         \\
R-squared           &       0.446         &       0.899         &       0.443         &       0.897         &       0.655         &       0.902         &       0.655         &       0.902         \\
Observations        &         406         &         212         &         406         &         212         &         406         &         214         &         406         &         214         \\

%% file: tableA1A.tex
Position on economic issues&       1.234\sym{***}&       1.202\sym{***}&       1.204\sym{***}&       0.885\sym{***}&                     &                     &       0.029         &       0.032         \\
                    &     (0.086)         &     (0.094)         &     (0.094)         &     (0.086)         &                     &                     &     (0.041)         &     (0.036)         \\
Position on economic issues sq.&      -0.122\sym{***}&      -0.120\sym{***}&      -0.120\sym{***}&      -0.083\sym{***}&                     &                     &                     &                     \\
                    &     (0.009)         &     (0.010)         &     (0.010)         &     (0.008)         &                     &                     &                     &                     \\
Centrism on economic issues&                     &                     &                     &                     &       0.590\sym{***}&       0.386\sym{***}&                     &                     \\
                    &                     &                     &                     &                     &     (0.049)         &     (0.039)         &                     &                     \\
Joint significance  &       0.000         &       0.000         &       0.000         &       0.000         &                     &                     &                     &                     \\
Vertex              &       5.046         &       5.020         &       5.010         &       5.315         &                     &                     &                     &                     \\
R-squared           &       0.109         &       0.157         &       0.167         &       0.386         &       0.164         &       0.383         &       0.070         &       0.361         \\
Observations        &        3622         &        3622         &        3622         &        3598         &        3622         &        3598         &        3622         &        3598         \\

%% file: tableA1B.tex
Position on social values&       1.554\sym{***}&       1.530\sym{***}&       1.536\sym{***}&       1.267\sym{***}&                     &                     &       0.033         &       0.149\sym{***}\\
                    &     (0.055)         &     (0.057)         &     (0.056)         &     (0.065)         &                     &                     &     (0.030)         &     (0.049)         \\
Position on social values sq.&      -0.147\sym{***}&      -0.145\sym{***}&      -0.146\sym{***}&      -0.112\sym{***}&                     &                     &                     &                     \\
                    &     (0.005)         &     (0.005)         &     (0.005)         &     (0.006)         &                     &                     &                     &                     \\
Centrism on social values&                     &                     &                     &                     &       0.800\sym{***}&       0.625\sym{***}&                     &                     \\
                    &                     &                     &                     &                     &     (0.030)         &     (0.036)         &                     &                     \\
Joint significance  &       0.000         &       0.000         &       0.000         &       0.000         &                     &                     &                     &                     \\
Vertex              &       5.295         &       5.263         &       5.262         &       5.666         &                     &                     &                     &                     \\
R-squared           &       0.283         &       0.313         &       0.319         &       0.432         &       0.323         &       0.429         &       0.054         &       0.345         \\
Observations        &        3527         &        3527         &        3527         &        3510         &        3527         &        3510         &        3527         &        3510         \\

%% file: tableAcentrismalt.tex
Centrism on economic issues &       0.575\sym{***}&       1.093\sym{***}&                     &                     \\
                    &     (0.082)         &     (0.301)         &                     &                     \\
Centrism on social values&                     &                     &       0.649\sym{***}&       0.670\sym{***}\\
                    &                     &                     &     (0.060)         &     (0.152)         \\
R-squared           &       0.365         &       0.903         &       0.460         &       0.901         \\
Observations        &         406         &         212         &         406         &         214         \\

%% file: tableA2f.tex
Position on economic issues&       0.428\sym{***}&       0.378\sym{***}&       0.350\sym{***}&       0.498\sym{***}&                     &                     &                     &                     \\
                    &     (0.030)         &     (0.033)         &     (0.034)         &     (0.087)         &                     &                     &                     &                     \\
Position on economic issues sq.&      -0.043\sym{***}&      -0.037\sym{***}&      -0.034\sym{***}&      -0.053\sym{***}&                     &                     &                     &                     \\
                    &     (0.003)         &     (0.003)         &     (0.003)         &     (0.008)         &                     &                     &                     &                     \\
Position on social values&                     &                     &                     &                     &       0.492\sym{***}&       0.464\sym{***}&       0.460\sym{***}&       0.701\sym{***}\\
                    &                     &                     &                     &                     &     (0.029)         &     (0.029)         &     (0.029)         &     (0.079)         \\
Position on social values sq.&                     &                     &                     &                     &      -0.048\sym{***}&      -0.046\sym{***}&      -0.046\sym{***}&      -0.072\sym{***}\\
                    &                     &                     &                     &                     &     (0.003)         &     (0.003)         &     (0.003)         &     (0.006)         \\
Joint significance  &       0.000         &       0.000         &       0.000         &       0.000         &       0.000         &       0.000         &       0.000         &       0.000         \\
Vertex              &       4.997         &       5.073         &       5.119         &       4.679         &       5.097         &       5.012         &       5.015         &       4.875         \\
R-squared           &       0.160         &       0.336         &       0.413         &       0.788         &       0.256         &       0.374         &       0.476         &       0.779         \\
Observations        &        1101         &        1101         &        1101         &         960         &        1102         &        1102         &        1102         &         962         \\

%% file: tableA2A.tex
Position on economic issues&       1.461\sym{***}&       1.894\sym{***}&       1.894\sym{***}&       1.550\sym{***}&                     &                     &                     &                     \\
                    &     (0.204)         &     (0.148)         &     (0.155)         &     (0.216)         &                     &                     &                     &                     \\
Position on economic issues sq.&      -0.147\sym{***}&      -0.192\sym{***}&      -0.192\sym{***}&      -0.156\sym{***}&                     &                     &                     &                     \\
                    &     (0.021)         &     (0.016)         &     (0.017)         &     (0.022)         &                     &                     &                     &                     \\
Centrism on economic issues&                     &                     &                     &                     &       0.567\sym{***}&       0.843\sym{***}&       0.856\sym{***}&       0.631\sym{***}\\
                    &                     &                     &                     &                     &     (0.104)         &     (0.082)         &     (0.085)         &     (0.110)         \\
Joint significance  &       0.000         &       0.000         &       0.000         &       0.000         &                     &                     &                     &                     \\
Vertex              &       4.974         &       4.933         &       4.921         &       4.951         &                     &                     &                     &                     \\
R-squared           &       0.582         &       0.547         &       0.552         &       0.613         &       0.533         &       0.474         &       0.476         &       0.554         \\
Observations        &         272         &         393         &         370         &         245         &         272         &         393         &         370         &         246         \\

%% file: tableA2B.tex
Position on social values&       1.276\sym{***}&       1.725\sym{***}&       1.618\sym{***}&       1.362\sym{***}&                     &                     &                     &                     \\
                    &     (0.144)         &     (0.096)         &     (0.102)         &     (0.154)         &                     &                     &                     &                     \\
Position on social values sq.&      -0.121\sym{***}&      -0.164\sym{***}&      -0.155\sym{***}&      -0.129\sym{***}&                     &                     &                     &                     \\
                    &     (0.014)         &     (0.009)         &     (0.010)         &     (0.015)         &                     &                     &                     &                     \\
Centrism on social values&                     &                     &                     &                     &       0.809\sym{***}&       1.009\sym{***}&       0.968\sym{***}&       0.837\sym{***}\\
                    &                     &                     &                     &                     &     (0.087)         &     (0.056)         &     (0.059)         &     (0.092)         \\
Joint significance  &       0.000         &       0.000         &       0.000         &       0.000         &                     &                     &                     &                     \\
Vertex              &       5.271         &       5.261         &       5.211         &       5.276         &                     &                     &                     &                     \\
R-squared           &       0.655         &       0.640         &       0.626         &       0.657         &       0.686         &       0.663         &       0.661         &       0.690         \\
Observations        &         267         &         392         &         369         &         239         &         267         &         392         &         369         &         240         \\

%% file: tableA3A.tex
Position on economic issues&       1.954\sym{***}&       1.962\sym{***}&       1.949\sym{***}&       1.962\sym{***}&                     &                     &       0.120\sym{**} &       0.101\sym{**} \\
                    &     (0.155)         &     (0.148)         &     (0.147)         &     (0.145)         &                     &                     &     (0.055)         &     (0.041)         \\
Position on economic issues sq.&      -0.197\sym{***}&      -0.197\sym{***}&      -0.195\sym{***}&      -0.198\sym{***}&                     &                     &                     &                     \\
                    &     (0.018)         &     (0.017)         &     (0.017)         &     (0.015)         &                     &                     &                     &                     \\
Centrism on economic issues&                     &                     &                     &                     &       0.866\sym{***}&       0.851\sym{***}&                     &                     \\
                    &                     &                     &                     &                     &     (0.091)         &     (0.080)         &                     &                     \\
Joint significance  &       0.000         &       0.000         &       0.000         &       0.000         &                     &                     &                     &                     \\
Vertex              &       4.965         &       4.984         &       4.989         &       4.958         &                     &                     &                     &                     \\
Observations        &         327         &         327         &         327         &         327         &         327         &         327         &         327         &         327         \\
R-squared           &       0.363         &       0.363         &       0.363         &       0.363         &       0.262         &       0.263         &       0.017         &       0.018         \\
KP F statistic (first stage)&     496.408         &      30.850         &      23.045         &      12.604         &      21.746         &      15.063         &      71.683         &      39.726         \\

%% file: tableA3B.tex
Position on social values&       1.861\sym{***}&       1.897\sym{***}&       1.862\sym{***}&       1.838\sym{***}&                     &                     &       0.017         &       0.022         \\
                    &     (0.126)         &     (0.121)         &     (0.119)         &     (0.098)         &                     &                     &     (0.039)         &     (0.031)         \\
Position on social values sq.&      -0.180\sym{***}&      -0.183\sym{***}&      -0.179\sym{***}&      -0.177\sym{***}&                     &                     &                     &                     \\
                    &     (0.012)         &     (0.011)         &     (0.011)         &     (0.009)         &                     &                     &                     &                     \\
Centrism on social values&                     &                     &                     &                     &       1.078\sym{***}&       1.081\sym{***}&                     &                     \\
                    &                     &                     &                     &                     &     (0.062)         &     (0.052)         &                     &                     \\
Joint significance  &       0.000         &       0.000         &       0.000         &       0.000         &                     &                     &                     &                     \\
Vertex              &       5.183         &       5.187         &       5.189         &       5.206         &                     &                     &                     &                     \\
Observations        &         328         &         328         &         328         &         328         &         328         &         328         &         328         &         328         \\
R-squared           &       0.527         &       0.526         &       0.527         &       0.527         &       0.576         &       0.576         &       0.001         &       0.001         \\
KP F statistic (first stage)&     320.928         &      26.359         &      20.684         &      10.856         &      20.247         &      11.661         &      74.456         &      43.343         \\

%% file: tableA3aa.tex
Centrism on economic issuesn (t)&                     &                     &       0.795\sym{***}&       0.823\sym{**} &                     &                     &                     &                     \\
                    &                     &                     &     (0.165)         &     (0.367)         &                     &                     &                     &                     \\
Centrism on economic issues (t-1)&       0.726\sym{***}&      -0.313         &       0.069         &      -0.115         &                     &                     &                     &                     \\
                    &     (0.088)         &     (0.238)         &     (0.158)         &     (0.228)         &                     &                     &                     &                     \\
Centrism on social values (t)&                     &                     &                     &                     &                     &                     &       1.048\sym{***}&       0.693\sym{***}\\
                    &                     &                     &                     &                     &                     &                     &     (0.124)         &     (0.168)         \\
Centrism on social values (t-1)&                     &                     &                     &                     &       0.886\sym{***}&      -0.126         &       0.007         &      -0.143         \\
                    &                     &                     &                     &                     &     (0.056)         &     (0.266)         &     (0.117)         &     (0.267)         \\
R-squared           &       0.378         &       0.900         &       0.429         &       0.910         &       0.563         &       0.896         &       0.693         &       0.908         \\
Observations        &         327         &         178         &         327         &         178         &         328         &         180         &         328         &         180         \\

%% file: table2altGDP.tex
Centrism on economic issues&       0.794\sym{***}&       0.693\sym{*}  &       1.928\sym{***}&                     &                     &                     \\
                    &     (0.098)         &     (0.369)         &     (0.458)         &                     &                     &                     \\
Centrism $\times$ GDP growth variance&       0.120         &      -0.015         &      -0.815\sym{*}  &                     &                     &                     \\
                    &     (0.136)         &     (0.633)         &     (0.421)         &                     &                     &                     \\
\multirow{1}{*}{\shortstack{Centrism $\times$ GDP growth var $\times$ not in govt }}&                     &       0.145         &       1.059\sym{**} &                     &                     &                     \\
                    &                     &     (0.647)         &     (0.427)         &                     &                     &                     \\
Centrism $\times$ not in government&                     &       0.106         &      -1.173\sym{***}&                     &                     &                     \\
                    &                     &     (0.392)         &     (0.434)         &                     &                     &                     \\
Centrism on social values&                     &                     &                     &       0.917\sym{***}&       1.087\sym{***}&       0.251         \\
                    &                     &                     &                     &     (0.073)         &     (0.160)         &     (0.625)         \\
Centrism $\times$ GDP growth variance&                     &                     &                     &   0.172\sym{\dagger}&       0.288         &      -0.085         \\
                    &                     &                     &                     &     (0.108)         &     (0.264)         &     (0.298)         \\
Centrism $\times$ GDP growth var $\times$ not in govt &                     &                     &                     &                     &      -0.148         &       0.166         \\
                    &                     &                     &                     &                     &     (0.292)         &     (0.335)         \\
Centrism $\times$ not in govt&                     &                     &                     &                     &      -0.199         &       0.431         \\
                    &                     &                     &                     &                     &     (0.184)         &     (0.642)         \\
GDP growth variance $\times$ not in government&                     &      -0.387         &      -2.937\sym{**} &                     &      -0.071         &      -0.321         \\
                    &                     &     (2.025)         &     (1.402)         &                     &     (0.770)         &     (0.931)         \\
Party not in government&                     &      -0.407         &       3.625\sym{**} &                     &       0.395         &      -0.896         \\
                    &                     &     (1.280)         &     (1.518)         &                     &     (0.456)         &     (1.839)         \\
R-squared           &       0.438         &       0.439         &       0.908         &       0.657         &       0.670         &       0.904         \\
Observations        &         349         &         349         &         194         &         348         &         348         &         196         \\

%% file: tableA_singleissuealt.tex
Position on economic issues&       1.421\sym{**} &       0.503\sym{***}&       0.808         &      -0.197\sym{**} \\
                    &     (0.589)         &     (0.090)         &     (0.578)         &     (0.086)         \\
Position on economic issues sq.&      -0.148\sym{***}&      -0.054\sym{***}&  -0.072\sym{\dagger}&       0.024\sym{***}\\
                    &     (0.050)         &     (0.008)         &     (0.048)         &     (0.008)         \\
Position on social values&      -0.383         &  -0.099\sym{\dagger}&       0.384         &       0.733\sym{***}\\
                    &     (0.449)         &     (0.061)         &     (0.418)         &     (0.076)         \\
Position on social values sq.&       0.035         &       0.012\sym{**} &      -0.057\sym{*}  &      -0.074\sym{***}\\
                    &     (0.040)         &     (0.006)         &     (0.030)         &     (0.006)         \\
Joint significance  &       0.678         &       0.101         &       0.321         &       0.003         \\
R-squared           &       0.896         &       0.791         &       0.897         &       0.785         \\
Observations        &         212         &         960         &         214         &         962         \\

%% file: tableA5.tex
\begin{table}[htbp]\centering \caption{Centrism across policy dimensions: simple correlations\label{tableA5}}\footnotesize
\begin{tabular}{l  c  c  c  c  c  c  c  c  c  c  c  c  c  c  c  c  c  c  c  c }\hline\hline
\multicolumn{1}{c}{Variables}  &(1) &(2) &(3) &(4) &(5) &(6) &(7) &(8) &(9) &(10)&(11)&(12)&(13)&(14) &(15) &(16)&(17)&(18) &(19)&(20)\\ \hline
Economic issues&1.000\\
 &\\
Social values&0.256&1.000\\
&(0.000) &\\
Immigration&0.330&0.686&1.000\\
&(0.000)&(0.000) &\\
Multiculturalism&0.220&0.689&0.852&1.000\\
&(0.000)&(0.000)&(0.000) &\\
Redistribution&0.817&0.223&0.275&0.183&1.000\\
&(0.000)&(0.000)&(0.000)&(0.000) &\\
Environment&0.358&0.503&0.519&0.482&0.368&1.000\\
&(0.000)&(0.000)&(0.000)&(0.000)&(0.000) &\\
Spending vs taxes&0.826&0.197&0.264&0.162&0.843&0.397&1.000\\
&(0.000)&(0.000)&(0.000)&(0.000)&(0.000)&(0.000) &\\
Deregulation&0.866&0.238&0.288&0.176&0.805&0.353&0.827&1.000\\
&(0.000)&(0.000)&(0.000)&(0.000)&(0.000)&(0.000)&(0.000) &\\
Econ intervention&0.858&0.254&0.289&0.217&0.852&0.360&0.835&0.889&1.000\\
&(0.000)&(0.000)&(0.000)&(0.000)&(0.000)&(0.000)&(0.000)&(0.000) &\\
Civil liberties&0.369&0.762&0.775&0.770&0.325&0.550&0.302&0.343&0.344&1.000\\
&(0.000)&(0.000)&(0.000)&(0.000)&(0.000)&(0.000)&(0.000)&(0.000)&(0.000) &\\
Lifestyle&0.247&0.815&0.528&0.547&0.289&0.448&0.257&0.276&0.321&0.648&1.000\\
&(0.000)&(0.000)&(0.000)&(0.000)&(0.000)&(0.000)&(0.000)&(0.000)&(0.000)&(0.000) &\\
Religion in politics&0.403&0.579&0.261&0.291&0.469&0.349&0.457&0.411&0.448&0.461&0.679&1.000\\
&(0.000)&(0.000)&(0.000)&(0.000)&(0.000)&(0.000)&(0.000)&(0.000)&(0.000)&(0.000)&(0.000) &\\
Minority rights&0.221&0.625&0.668&0.714&0.222&0.433&0.178&0.179&0.214&0.685&0.539&0.398&1.000\\
&(0.000)&(0.000)&(0.000)&(0.000)&(0.000)&(0.000)&(0.000)&(0.000)&(0.000)&(0.000)&(0.000)&(0.000) &\\
Nationalism&0.239&0.760&0.757&0.804&0.141&0.484&0.123&0.171&0.177&0.773&0.637&0.414&0.752&1.000\\
&(0.000)&(0.000)&(0.000)&(0.000)&(0.001)&(0.000)&(0.004)&(0.000)&(0.000)&(0.000)&(0.000)&(0.000)&(0.000) &\\
Urban/Rural&0.068&0.254&0.082&0.106&0.063&0.141&0.037&0.133&0.117&0.146&0.206&0.253&0.090&0.268&1.000\\
&(0.058)&(0.000)&(0.023)&(0.003)&(0.081)&(0.000)&(0.296)&(0.000)&(0.006)&(0.000)&(0.000)&(0.000)&(0.012)&(0.000) &\\
Protectionism&0.422&0.385&0.506&0.419&0.231&0.267&0.216&0.376&0.367&0.469&0.294&0.221&0.302&0.467&0.359&1.000\\
&(0.000)&(0.000)&(0.000)&(0.000)&(0.000)&(0.000)&(0.000)&(0.000)&(0.000)&(0.000)&(0.000)&(0.000)&(0.000)&(0.000)&(0.000) &\\
Decentralization&0.122&0.226&0.195&0.267&0.088&0.192&0.070&0.120&0.123&0.260&0.174&0.171&0.418&0.413&0.089&0.224&1.000\\
&(0.001)&(0.000)&(0.000)&(0.000)&(0.014)&(0.000)&(0.051)&(0.001)&(0.004)&(0.000)&(0.000)&(0.005)&(0.000)&(0.000)&(0.013)&(0.000) &\\
Anti-Islam&0.063&0.202&0.004&0.075&0.087&0.166&0.090&0.040&0.087&0.183&0.312&0.198&0.247&0.224&0.104&0.055&0.051&1.000\\
&(0.305)&(0.001)&(0.952)&(0.217)&(0.151)&(0.006)&(0.141)&(0.510)&(0.155)&(0.002)&(0.000)&(0.001)&(0.000)&(0.000)&(0.087)&(0.367)&(0.410) &\\
Anti-elite&0.180&0.241&0.395&0.386&0.093&0.253&0.089&0.176&0.141&0.374&0.095&0.018&0.334&0.344&0.078&0.365&0.348&-0.006&1.000\\
&(0.000)&(0.000)&(0.000)&(0.000)&(0.123)&(0.000)&(0.142)&(0.003)&(0.019)&(0.000)&(0.116)&(0.765)&(0.000)&(0.000)&(0.195)&(0.000)&(0.000)&(0.922) &\\
European integration&-0.058&-0.036&-0.032&-0.036&-0.101&-0.025&-0.058&-0.020&-0.085&0.031&0.013&-0.053&0.045&0.064&0.108&0.150&-0.005&0.126&0.001&1.000\\
&(0.056)&(0.234)&(0.281)&(0.234)&(0.005)&(0.483)&(0.107)&(0.581)&(0.047)&(0.387)&(0.725)&(0.379)&(0.178)&(0.139)&(0.002)&(0.013)&(0.890)&(0.038)&(0.981)\\
\hline \hline 
\multicolumn{21}{p{18cm}}{\footnotesizes{\textbf{Notes:}  The p-values are reported in parentheses.} } \\
 \end{tabular}
\end{table}

%% file: tableA6a.tex
Centrism on economic issues&       0.544\sym{***}&   0.620\sym{\dagger}&       0.655\sym{***}&       1.243\sym{**} &                     &                     &                     &                     \\
                    &     (0.153)         &     (0.363)         &     (0.158)         &     (0.569)         &                     &                     &                     &                     \\
Centrism $\times$ GDP growth&       0.105\sym{**} &       0.082         &                     &                     &                     &                     &                     &                     \\
                    &     (0.045)         &     (0.059)         &                     &                     &                     &                     &                     &                     \\
Centrism $\times$ years with GDP growth$<0$&                     &                     &       1.003         &      -2.114         &                     &                     &                     &                     \\
                    &                     &                     &     (0.926)         &     (2.327)         &                     &                     &                     &                     \\
Centrism on social values&                     &                     &                     &                     &       0.844\sym{***}&       0.759\sym{***}&       0.768\sym{***}&   0.967\sym{\dagger}\\
                    &                     &                     &                     &                     &     (0.106)         &     (0.120)         &     (0.126)         &     (0.598)         \\
Centrism $\times$ GDP growth&                     &                     &                     &                     &   0.052\sym{\dagger}&      -0.024         &                     &                     \\
                    &                     &                     &                     &                     &     (0.032)         &     (0.033)         &                     &                     \\
Centrism $\times$ years with GDP growth $<0$&                     &                     &                     &                     &                     &                     &       1.381\sym{**} &      -1.719         \\
                    &                     &                     &                     &                     &                     &                     &     (0.658)         &     (3.524)         \\
R-squared           &       0.448         &       0.898         &       0.444         &       0.897         &       0.657         &       0.902         &       0.659         &       0.902         \\
Observations        &         406         &         212         &         406         &         212         &         406         &         214         &         406         &         214         \\

%% file: tableA6b.tex
Centrism on economic issues&       0.756\sym{***}&       0.835\sym{***}&       0.792\sym{***}&       0.792\sym{***}&                     &                     &                     &                     \\
                    &     (0.274)         &     (0.082)         &     (0.079)         &     (0.079)         &                     &                     &                     &                     \\
Centrism $\times$ years with banking crisis&       1.956         &                     &                     &                     &                     &                     &                     &                     \\
                    &     (9.854)         &                     &                     &                     &                     &                     &                     &                     \\
Centrism $\times$ years with  currency crisis&                     &      -2.245         &                     &                     &                     &                     &                     &                     \\
                    &                     &     (4.499)         &                     &                     &                     &                     &                     &                     \\
Centrism $\times$ years with debt crisis&                     &                     &       4.320         &                     &                     &                     &                     &                     \\
                    &                     &                     &     (9.188)         &                     &                     &                     &                     &                     \\
Centrism $\times$ years with debt restructuring&                     &                     &                     &       4.320         &                     &                     &                     &                     \\
                    &                     &                     &                     &     (9.188)         &                     &                     &                     &                     \\
Centrism on social values&                     &                     &                     &                     &       1.056\sym{***}&       1.012\sym{***}&       0.953\sym{***}&       0.953\sym{***}\\
                    &                     &                     &                     &                     &     (0.149)         &     (0.060)         &     (0.057)         &     (0.057)         \\
Centrism $\times$ years with banking crisis&                     &                     &                     &                     &      -2.779         &                     &                     &                     \\
                    &                     &                     &                     &                     &     (4.413)         &                     &                     &                     \\
Centrism $\times$ years with  currency crisis&                     &                     &                     &                     &                     &      -2.699         &                     &                     \\
                    &                     &                     &                     &                     &                     &     (2.029)         &                     &                     \\
Centrism $\times$ years with debt crisis&                     &                     &                     &                     &                     &                     &       5.711         &                     \\
                    &                     &                     &                     &                     &                     &                     &     (5.025)         &                     \\
Centrism $\times$ years with debt restructuring&                     &                     &                     &                     &                     &                     &                     &       5.711         \\
                    &                     &                     &                     &                     &                     &                     &                     &     (5.025)         \\
R-squared           &       0.442         &       0.442         &       0.442         &       0.442         &       0.656         &       0.657         &       0.657         &       0.657         \\
Observations        &         404         &         404         &         404         &         404         &         404         &         404         &         404         &         404         \\